\documentclass[times,twocolumn,final]{elsarticle}

\usepackage{medima}
\usepackage{framed,multirow}
\usepackage{amssymb}
\usepackage{latexsym}
\usepackage{amsmath}
\usepackage{url}
\usepackage{siunitx}
\usepackage{nccmath}
\usepackage{empheq}
\usepackage{booktabs}
\usepackage{balance}

\journal{Medical Image Analysis}
\begin{document}
\verso{B. Billot \textit{et~al.}}

\begin{frontmatter}

\title{SynthSeg: Segmentation of brain MRI scans of any contrast and resolution without retraining}

\author[1]{Benjamin Billot\corref{cor1}}
\author[2]{Douglas N. Greve}
\author[3]{Oula Puonti}
\author[3,4]{Axel Thielscher}
\author[2,4]{Koen Van Leemput}
\author[2,5,6]{Bruce Fischl}
\author[2,5]{Adrian V. Dalca}
\author[1,2,5]{Juan Eugenio Iglesias}
\author[]{for the ADNI \fnref{footnote}}
\fntext[footnote]{Data used in this article are partly from the Alzheimer's Disease Neuroimaging Initiative database (http://adni.loni.usc.edu). Investigators in the ADNI contributed to the design and implementation of ADNI and/or provided data but did not participate in analysis of this report. A complete listing of investigators at: adni.loni.usc.edu/wp-content/ADNI\_Acknowledgement\_List.pdf}

\cortext[cor1]{Corresponding author: \emph{email}: benjamin.billot.18@ucl.ac.uk}
\address[1]{Centre for Medical Image Computing, University College London, UK}
\address[2]{Martinos Center for Biomedical Imaging, Department of Radiology, Massachusetts General Hospital and Harvard Medical School, USA}
\address[3]{Danish Research Centre for Magnetic Resonance, Centre for Functional and Diagnostic Imaging and Research, Copenhagen University Hospital, Denmark}
\address[4]{Department of Health Technology, Technical University of Denmark}
\address[5]{Computer Science and Artificial Intelligence Laboratory, Massachusetts Institute of Technology, USA}
\address[6]{Program in Health Sciences and Technology, Massachusetts Institute of Technology, USA}


\begin{abstract}
Despite advances in data augmentation and transfer learning, convolutional neural networks (CNNs) difficultly generalise to unseen domains. When segmenting brain scans, CNNs are highly sensitive to changes in resolution and contrast: even within the same MRI modality, performance can decrease across datasets. Here we introduce SynthSeg, the first segmentation CNN robust against changes in contrast and resolution. SynthSeg is trained with synthetic data sampled from a generative model conditioned on segmentations. Crucially, we adopt a \textit{domain randomisation} strategy where we fully randomise the contrast and resolution of the synthetic training data. Consequently, SynthSeg can segment real scans from a wide range of target domains without retraining or fine-tuning, which enables straightforward analysis of huge amounts of heterogeneous clinical data. Because SynthSeg only requires segmentations to be trained (no images), it can learn from labels obtained by automated methods on diverse populations (e.g., ageing and diseased), thus achieving robustness to a wide range of morphological variability. We demonstrate SynthSeg on 5,000 scans of six modalities (including CT) and ten resolutions, where it exhibits unparalleled generalisation compared with supervised CNNs, state-of-the-art domain adaptation, and Bayesian segmentation. Finally, we demonstrate the generalisability of SynthSeg by applying it to cardiac MRI and CT~scans.
\end{abstract}


\end{frontmatter}


\section{Introduction}
\label{sec:introduction}

\subsection{Motivation}

Segmentation of brain scans is of paramount importance in neuroimaging, as it enables volumetric and shape analyses~\citep{hynd_corpus_1991}. Although manual delineation is considered the gold standard in segmentation, this procedure is tedious and costly, thus preventing the analysis of large datasets. Moreover, manual segmentation of brain scans requires expertise in neuroanatomy, which, even if available, suffers from severe inter- and intra-rater variability issues~\citep{warfield_simultaneous_2004}. For these reasons, automated segmentation methods have been proposed as a fast and reproducible alternative solution.

Most recent automated segmentation methods rely on convolutional neural networks (CNNs)~\citep{ronneberger_u-net_2015, milletari_v-net_2016, kamnitsas_efficient_2017}. These are widespread in research, where the abundance of high quality scans (i.e., at high isotropic resolution and with good contrasts between tissues) enables CNNs to obtain accurate 3D segmentations that can then be used in subsequent analyses such as connectivity study~\citep{muller_underconnected_2011}. 

However, supervised CNNs are far less employed in clinical settings, where physicians prefer 2D acquisitions with a sparse set of high-resolution slices, which enables faster inspection under time constraints. This leads to a huge variability in image orientation (axial, coronal, or sagittal), slice spacing, and in-plane resolution. Moreover, such 2D scans often use thick slices to increase the signal-to-noise ratio, thus introducing considerable partial voluming (PV). This effect arises when several tissue types are mixed within the same voxel, resulting in averaged intensities that are not necessarily representative of underlying tissues, often causing segmentation methods to underperform~\citep{van_leemput_unifying_2003}. Additionally, imaging protocols also span a huge diversity in sequences and modalities, each studying different tissue properties, and the resulting variations in intensity distributions drastically decrease the accuracy of supervised CNNs~\citep{chen_synergistic_2019}.

Overall, the lack of tools that can cope with the large variability in MR data hinders the adoption of quantitative morphometry in the clinic. Moreover, it precludes the analysis of vast amounts of clinical scans, currently left unexplored in picture archiving and communication systems (PACS) in hospitals around the world. The ability to derive morphometric measurements from these scans would enable neuroimaging studies with sample sizes in the millions, and thus much higher statistical power than current research studies. Therefore, there is a clear need for a fast, accurate, and reproducible automated method, for segmentation of brain scans of any contrast and resolution, and that can adapt to a wide range of populations.

\begin{figure}[!ht]
    \centering
    \includegraphics[width=0.97\columnwidth]{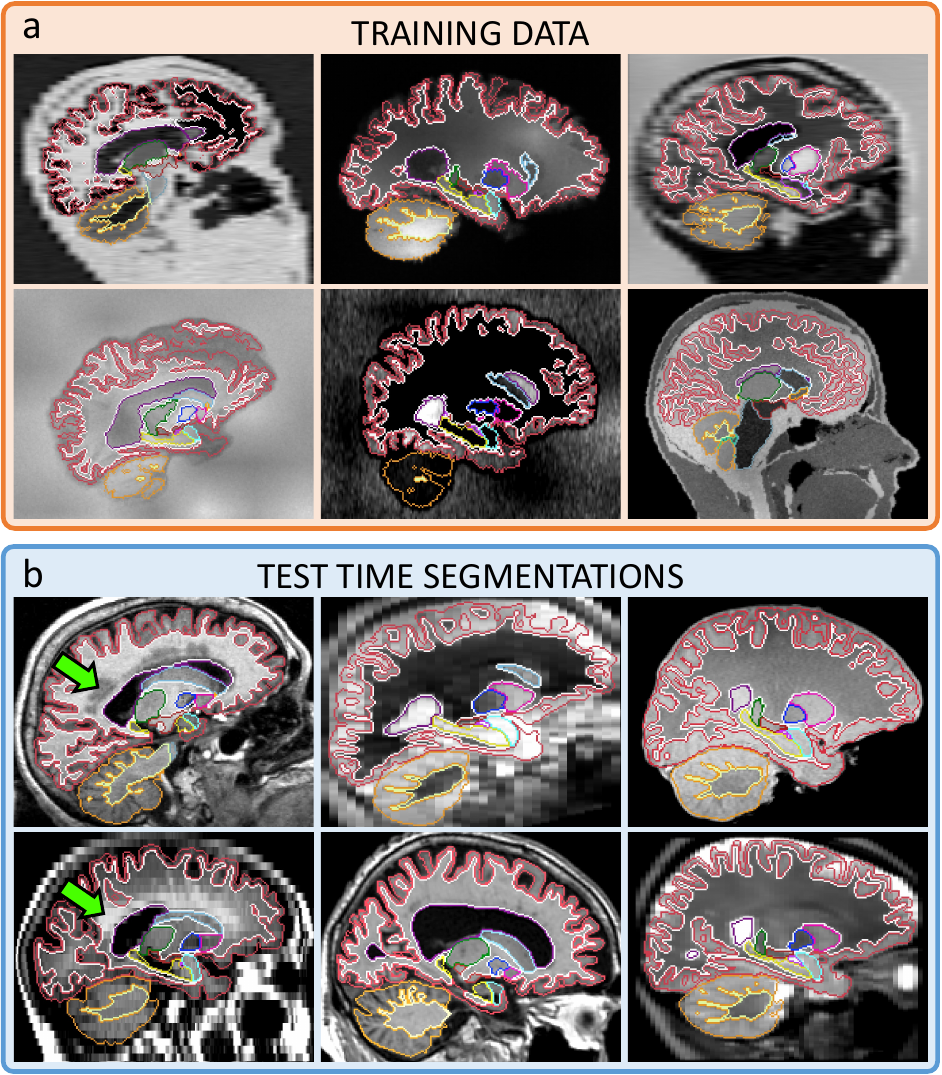}
    \caption{(a)~Representative samples of the synthetic 3D scans used to train SynthSeg for brain segmentation, and contours of the corresponding ground truth. (b)~Test-time segmentations for a variety of contrasts and resolutions, on subjects spanning a wide age range, some presenting large atrophy and white matter lesions (green arrows). All segmentations are obtained with the same network, without retraining or fine-tuning.}
    \label{fig:front_page}
\end{figure}

\subsection{Contributions}

In this article, we present SynthSeg, the first neural network to segment brain scans of a wide range of contrasts and resolutions, without having to be retrained or fine-tuned (Figure~\ref{fig:front_page}). Specifically, SynthSeg is trained with synthetic scans sampled on the fly from a generative model inspired by the Bayesian segmentation framework, and is thus never exposed to real scans during training. Our main contribution is the adoption of a domain \textit{randomisation strategy}~\citep{tobin_domain_2017}, where all the parameters of the generative model (including orientation, contrast, resolution, artefacts) are fully randomised. This exposes the network to vastly different examples at each mini-batch, and thus forces it to learn domain-independent features. Moreover, we apply a random subset of common preprocessing operations to each example (e.g., skull stripping, bias field correction), such that SynthSeg can segment scans with or without preprocessing. 

With this domain randomisation strategy, our method only needs to be trained once. This is a considerable improvement over supervised CNNs and domain adaptation strategies, which all need retraining or fine-tuning for each new contrast or resolution, thus hindering clinical applications. Moreover, training SynthSeg is greatly facilitated by the fact that it only requires a set of anatomical label maps to be trained (and no real images, since all training scans are synthetic). Furthermore, these maps can be obtained automatically (rather than manually), since the training scans are directly generated from their ground truths, and are thus perfectly aligned with them. This enables us to greatly improve the robustness of SynthSeg by including automated training maps from highly diverse populations.

Overall, SynthSeg yields almost the accuracy of supervised CNNs on their training domain, but unlike them, exhibits a remarkable generalisation ability. Indeed, SynthSeg consistently outperforms state-of-the-art domain adaptation strategies and Bayesian segmentation on all tested datasets. Moreover, we demonstrate the generalisability of SynthSeg by obtaining state-of-the-art results in cross-modality cardiac segmentation.

This work extends our recent articles on contrast-adaptiveness~\citep{billot_learning_2020} and PV simulation at a specific resolution~\citep{billot_partial_2020, iglesias_joint_2021}, by building, for the first time, robustness to both contrast and resolution without retraining. Our method is thoroughly evaluated in four new experiments. The code and trained model are available at https://github.com/BBillot/SynthSeg as well as in the widespread neuroimaging package FreeSurfer~\citep{fischl_freesurfer_2012}.

\section{Related works}

\textbf{Contrast-invariance in brain segmentation} has traditionally been addressed with Bayesian segmentation. This technique is based on a generative model, which combines an anatomical prior (often a statistical atlas) and an intensity likelihood (typically a Gaussian Mixture Model, GMM). Scans are then segmented by ``inverting'' this model with Bayesian inference~\citep{wells_multi-modal_1996,fischl_whole_2002}. Contrast-robustness is achieved by using an unsupervised likelihood model, with parameters estimated on each test scan~\citep{van_leemput_automated_1999,ashburner_unified_2005}. However, Bayesian segmentation requires approximately 15 minutes per scan~\citep{puonti_fast_2016}, which precludes its use in time-sensitive settings. Additionally, its accuracy is limited at low resolution (LR) by PV effects~\citep{choi_partial_1991}. Indeed, even if Bayesian methods can easily model PV~\citep{van_leemput_unifying_2003}, inferring high resolution (HR) segmentations from LR scans quickly becomes intractable, as it requires marginalising over all possible configurations of HR labels within each LR supervoxel. While simplifications can be made~\citep{van_leemput_unifying_2003}, PV-aware Bayesian segmentation may still be infeasible in clinical settings.

\textbf{Supervised CNNs} prevail in recent medical image segmentation~\citep{milletari_v-net_2016, kamnitsas_efficient_2017}, and are best represented by the UNet architecture~\citep{ronneberger_u-net_2015}. While these networks obtain fast and accurate results on their training domain, they do not generalise well to unseen contrasts~\citep{karani_lifelong_2018} and resolutions~\citep{ghafoorian_transfer_2017}, an issue known as the ``domain-gap'' problem~\citep{pan_survey_2010}. Therefore, such networks need to be retrained for any new combination of contrast and resolution, often requiring new costly labelled data. This problem can partly be ameliorated by training on multi-modality scans with modality dropout~\citep{havaei_hemis_2016}, which results in a network able to individually segment each training modality, but that still cannot be applied to unseen domains. 

\textbf{Data augmentation} improves the robustness of CNNs by applying simple spatial and intensity transforms to the training data~\citep{zhang_generalizing_2020}. While such transforms often relies on handcrafted (and thus suboptimal) parameters, recent semi-supervised methods, such as adversarial augmentation, explicitly optimise the augmentation parameters during training~\citep{zhang_deep_2017,chaitanya_semi-supervised_2019,chen_enhancing_2022}. Alternatively, contrastive learning methods have been proposed to leverage unsupervised data for improved generalisation ability~\citep{chaitanya_contrastive_2020,you_simcvd_2022}. Overall, although these techniques improve generalisation in intra-modality applications~\citep{zhao_data_2019}, they generally remain insufficient in cross-modality settings~\citep{karani_lifelong_2018}.

\textbf{Domain adaptation} explicitly seeks to bridge a given domain gap between a source domain with labelled data, and a specific target domain without labels. A first solution is to map both domains to a common latent space, where a classifier can be trained~\citep{kamnitsas_unsupervised_2017, dou_plug_2019, ganin_domain-adversarial_2017,you_incremental_2022}. In comparison, generative adaptation methods seek to match the source images to the target domain with image-to-image translation methods~\citep{sandfort_data_2019, huo_synseg-net_2019, zhang_translating_2018}. Since these approaches are complementary, recent methods propose to operate in both feature and image space, which leads to state-to-the-art results in cross-modality segmentation~\citep{chen_synergistic_2019, hoffman_cycada_2018}. In contrast, state-of-the-art results in intra-modality adaptation are obtained with test-time adaptation methods~\citep{karani_test-time_2021, he_autoencoder_2021}, which rely on light fine-tuning at test-time. More generally, even though domain adaptation alleviates the need for supervision in the target domain, it still needs retraining for each new domain.

\textbf{Synthetic training data} can be used to increase robustness by introducing surrogate domain variations, either generated with physics-based models~\citep{jog_pulse_2018}, or adversarial generative networks~\citep{frid_synthetic_2018,chartsias_multimodal_2018}, possibly conditioned on label maps for improved semantic content~\citep{mahmood_deep_2020, isola_image--image_2017}. These strategies enable to generate huge training datasets with perfect ground truth obtained by construction rather than human annotation~\citep{richter_playing_2016}. However, although generated images may look remarkably realistic, they still suffer from a ``reality gap''~\citep{jakobi_noise_1995}. In addition, these methods still require retraining for every new domain, and thus do not solve the lack of generalisation of neural networks. To the best of our knowledge, no current learning method can segment medical scans of any contrast and/or resolution without retraining.

\textbf{Domain randomisation} is a recent strategy that relies on physics-based generative models, which, unlike learning-based methods, offer full control over the generation process. Instead of handcrafting~\citep{jog_pulse_2018} or optimising~\citep{chen_enhancing_2022} this kind of generative model to match a specific domain, Domain Randomisation (DR) proposes to considerably enlarge the distribution of the synthetic data by \textit{fully randomising} the generation parameters~\citep{tobin_domain_2017}. This learning strategy is motivated by converging evidence that augmentation beyond realism leads to improved generalisation~\citep{bengio_deep_2011, zhao_data_2019}. If pushed to the extreme, DR yields highly unrealistic samples, in which case real images are encompassed within the landscape of the synthetic training data~\citep{tremblay_training_2018}. As a result, this approach seeks to bridge all domain gaps in a given semantic space, rather than solving this problem for each domain gap separately. So far, DR has been used to control robotic arms~\citep{tobin_domain_2017}, and for car detection in street views~\citep{tremblay_training_2018}. Here we combine DR with a generative model inspired by Bayesian segmentation, in order to achieve, for the first time, segmentation of brain MRI scans of a wide range of contrasts and resolutions without retraining.

\section{Methods}

\subsection{Generative model}
\label{sec:generative model}

\begin{figure}[t]
    \centering
    \includegraphics[width=\columnwidth]{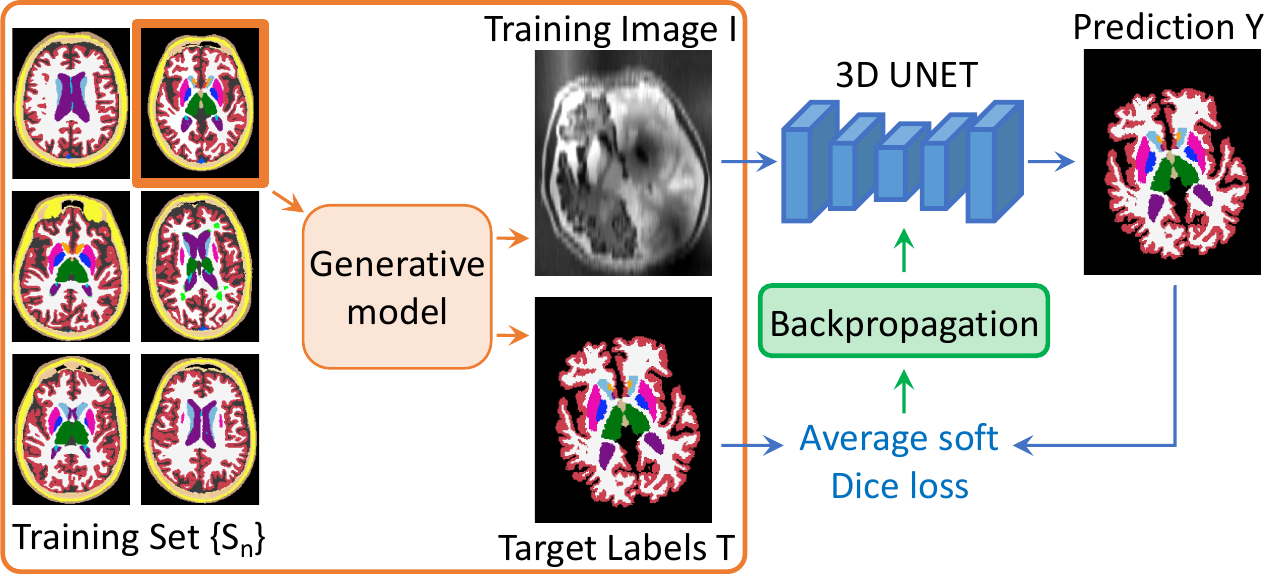}
    \caption{Overview of a training step. At each mini-batch, we randomly select a 3D label map from a training set $\{S_n\}$ and sample a pair $\{I, T\}$ from the generative model. The obtained image is then run through the network, and its prediction $Y$ is used to compute the average soft Dice loss, that is backpropagated to update the weights of the network.}
    \label{fig:overview}
\end{figure}

SynthSeg relies on a generative model from which we sample synthetic scans to train a segmentation network~\citep{billot_learning_2020, billot_partial_2020,iglesias_joint_2021}. Crucially, the training images are all generated on the fly with fully randomised parameters, such that the network is exposed to a different combination of contrast, resolution, morphology, artefacts, and noise at each mini-batch (Figure~\ref{fig:overview}). Here we describe the generative model, which is illustrated in Figure~\ref{fig:generative model} and exemplified in Supplement~1.

\vspace{2mm}
\subsubsection{Label map selection and spatial augmentation}
\label{sec:spatial augm}

The proposed generative model assumes the availability of $N$ training label maps $\{S_n\}_{n=1}^{N}$ defined over discrete spatial coordinates $(x,y,z)$ at high resolution $r_{HR}$. Let all label maps take their values from a set of $K$ labels: $S_n(x,y,z) \in \{1,...,K\}$. We emphasise that these training label maps can be obtained manually, automatically (by segmenting brain scans with an automated method), or even can be a combination thereof -- as long as they share the same labelling convention.

\begin{figure*}[t]
    \centering
    \includegraphics[width=0.9\textwidth]{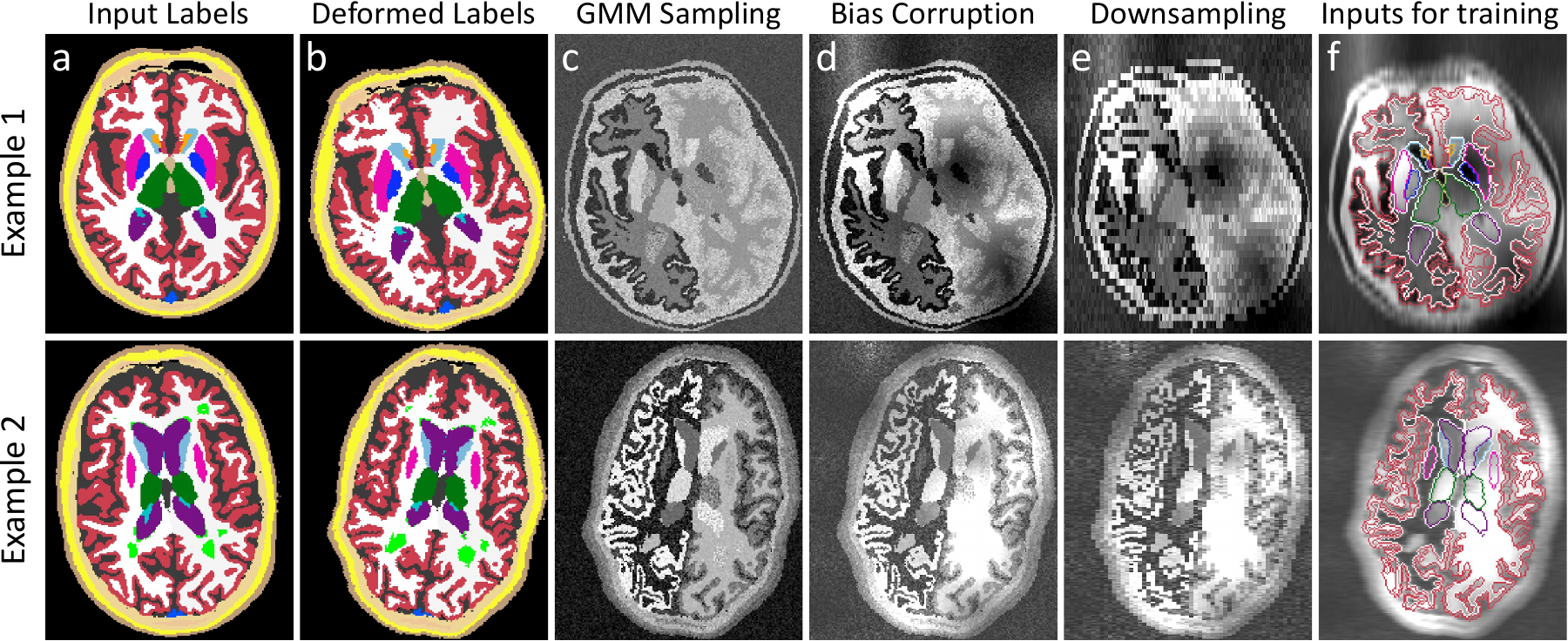}
    \caption{Intermediate steps of the generative model: (a) we randomly select an input label map from the training set, which we (b) spatially augment in 3D. (c) A first synthetic image is obtained by sampling a GMM at HR with randomised parameters. (d) The result is then corrupted with a bias field and further intensity augmentation. (e) Slice spacing and thickness are simulated by successively blurring and downsampling at random LR. (f) The training inputs are obtained by resampling the image to HR, and removing the labels we do not wish to segment (e.g., extra-cerebral regions).}
    \label{fig:generative model}
\end{figure*}

The generative process starts by randomly selecting a segmentation $S_i$ from the training dataset (Figure~\ref{fig:generative model}a). In order to increase the variability of the available segmentations, $S_i$ is deformed with a random spatial transform $\phi$, which is the composition of an affine and a non-linear transform.

The affine transformation $\phi_{\mathrm{aff}}$ is the composition of three rotations ($\theta_x$, $\theta_y$, $\theta_z$), three scalings ($s_x$, $s_y$, $s_z$), three shearings ($sh_x$, $sh_y$, $sh_z$), and three translations ($t_x$, $t_y$, $t_z$), whose parameters are sampled from uniform distributions:
\begin{align}
& \theta_x, \theta_y, \theta_z \sim \mathcal{U}(a_{rot},b_{rot}), \\
& s_x, s_y, s_z \sim \mathcal{U}(a_{sc},b_{sc}),  \\
& sh_x, sh_y, sh_z \sim \mathcal{U}(a_{sh},b_{sh}), \\
& t_x, t_y, t_z \sim \mathcal{U}(a_{tr},b_{tr}), \\
& \phi_{\mathrm{aff}} = \text{Aff}(\theta_x, \theta_y, \theta_z, s_x, s_y, s_z, sh_x, sh_y, sh_z, t_x, t_y, t_z),
\end{align}
where $a_{rot}, b_{rot}, a_{sc}, b_{sc}, a_{sh}, b_{sh}, a_{tr}, b_{tr}$ are the predefined bounds of the uniform distributions, and $\mathrm{Aff}(\cdot)$ refers to the composition of the aforementioned affine transforms.

The non-linear component $\phi_{\mathrm{nonlin}}$ is a diffeomorphic transform obtained as follows. First, we sample a small vector field of size $10\times10\times10\times3$ from a zero-mean Gaussian distribution of standard deviation $\sigma_{\mathrm{SVF}}$ drawn from $\mathcal{U}(0, b_{\mathrm{nonlin}})$. This field is then upsampled to full image size with trilinear interpolation to obtain a stationary velocity field (SVF). Finally, we integrate this SVF with a scale-and-square approach~\citep{arsigny_log-euclidean_2006} to yield a diffeomorphic deformation field that does not produce holes or foldings:

\begin{align}
    \sigma_{SVF} & \sim \mathcal{U}(0, b_{\mathrm{nonlin}}), \\
    \text{SVF}' & \sim \mathcal{N}_{10\times10\times10\times3}(0, \sigma_{\mathrm{SVF}}), \\
    \text{SVF} &= \mathrm{Resample}(\text{SVF}'; r_{HR}), \\
    \phi_{\mathrm{nonlin}} &= \mathrm{Integrate}(\text{SVF}).
\end{align}

Finally, we obtain an augmented map $L$ by applying $\phi$ to $S_i$ using nearest neighbour interpolation (Figure~\ref{fig:generative model}b):

\begin{equation}
L = S_i \circ \phi = S_i \circ (\phi_{\mathrm{aff}} \circ \phi_{\mathrm{nonlin}}).
\label{eq:spatial_augm}
\end{equation}

\subsubsection{Initial HR synthetic image}
\label{sec:hr image}

After deforming the input segmentation, we generate an initial synthetic scan $G$ at HR by sampling a GMM conditioned on $L$ (Figure~\ref{fig:generative model}c). For convenience, we regroup all the means and standard deviations of the GMM in $M_G = \{ \mu_k\}_{1\leq k \leq K}$ and $\Sigma_G = \{ \sigma_k\}_{1\leq k \leq K}$ respectively. Crucially, in order to randomise the contrast of $G$, all the parameters in $M_G$ and $\Sigma_G$ are sampled at each mini-batch from uniform distributions of range $\{a_\mu, b_\mu\}$ and $\{a_\sigma, b_\sigma\}$, respectively. We highlight that $\Sigma_G$ jointly models tissue heterogeneities as well as the thermal noise of the scanner. $G$ is then formed by independently sampling at each location $(x,y,z)$ the distribution indexed by $L(x,y,z)$:

\begin{align}
\mu_k \sim & \mathcal{U}(a_\mu, b_\mu), \\
\sigma_k \sim & \mathcal{U}(a_\sigma, b_\sigma), \\
G(x,y,z) \sim & \mathcal{N}(\mu_{L(x,y,z)},\sigma^2_{L(x,y,z)}).
\end{align}

\subsubsection{Bias field and intensity augmentation}
\label{section:intensity_augm}

We then simulate bias field artefacts to make SynthSeg robust to such effects. We sample a small volume of shape $4^3$ from a zero-mean Gaussian distribution of random standard deviation $\sigma_B$. We then upsample this small volume to full image size, and take the voxel-wise exponential to obtain a smooth and non-negative field $B$. Finally, we multiply $G$ by $B$ to obtain a biased image $G_B$ (Figure~\ref{fig:generative model}d), where the previous exponential ensures that division and multiplication by the same factor are equally likely~\citep{van_leemput_automated_1999, ashburner_unified_2005}:

\begin{align}
\sigma_{B} & \sim \mathcal{U}(0, b_B), \\
B' &\sim \mathcal{N}_{4 \times 4 \times 4 }(0,\sigma^2_B), \\
B &= \text{Upsample}(B'), \\
G_B(x,y,z) &= G(x,y,z) \times \exp [B(x,y,z)].
\end{align}

Then, a final HR image $I_{HR}$ is produced by rescaling $G_B$ between 0 and 1, and applying a random Gamma transform (voxel-wise exponentiation) to further augment the intensity distribution of the synthetic scans. This transform enables us to skew the distribution while leaving intensities in the $[0,1]$ interval. In practice, the exponent is sampled in the logarithmic domain from a zero-mean Gaussian distribution of standard deviation $\sigma_\gamma$. As a result, $I_{HR}$ is given by:

\begin{align}
\gamma \sim & \mathcal{N}(0, \sigma_\gamma^2), \\
I_{HR}(x,y,z) = & \left( \frac{G(x,y,z) -\min_{x,y,z} G}{\max_{x,y,z} G - \min_{x,y,z} G} \right) ^{\exp(\gamma)}.
\end{align}

\subsubsection{Simulation of resolution variability}

In order to make the network robust against changes in resolution, we now model differences in acquisition direction (i.e., axial, coronal, sagittal), slice spacing, and slice thickness. After randomly selecting a direction, the slice spacing $r_{spac}$ and slice thickness $r_{thick}$ are respectively drawn from $\mathcal{U}(r_{HR}, b_{res})$ and $\mathcal{U}(r_{HR}, r_{spac})$. Note that $r_{thick}$ is bound by $r_{spac}$ as slices very rarely overlap in practice.

Once all resolution parameters have been sampled, we first simulate slice thickness by blurring $I_{HR}$ into $I_\sigma$ with a Gaussian kernel that approximates the real slice excitation profile. Specifically, its standard deviation $\sigma_{thick}$ is designed to divide the power of the HR signal by 10 at the cut-off frequency~\citep{billot_partial_2020}. Moreover, $\sigma_{thick}$ is multiplied by a random coefficient $\alpha$ to introduce small deviations from the nominal thickness, and to mitigate the Gaussian assumption.

Slice spacing is then modelled by downsampling $I_\sigma$ to $I_{LR}$ at the prescribed low resolution $r_{spac}$ with trilinear interpolation (Figure~\ref{fig:generative model}e)~\citep{van_leemput_unifying_2003}. Finally, $I_{LR}$ is upsampled back to $r_{HR}$ (typically \SI{1}{\milli\meter}), such that the CNN is trained to produce crisp HR segmentations, regardless of the simulated resolution. This process can be summarised as:

\begin{align}
r_{spac} &\sim \mathcal{U}(r_{HR}, b_{res}), \\
r_{thick} &\sim \mathcal{U}(r_{HR}, r_{spac}), \\
\alpha &\sim \mathcal{U}(a_\alpha, b_\alpha), \\
\sigma_{thick} &= 2 \alpha \log(10) (2\pi)^{-1} r_{thick} / r_{HR} ,  \\
I_\sigma &= I_{HR} \ast \mathcal{N}(0, \sigma_{thick}), \\
I_{LR} &= \text{Resample} \left( I_\sigma ; r_{spac} \right), \\
I &= \text{Resample} \left( I_{LR} ; r_{HR} \right).
\end{align}

\subsubsection{Model output and segmentation target}

At each training step, our method produces two volumes: an image $I$ sampled from the generative model, and its segmentation target $T$. The latter is obtained by taking the deformed map $L$ in~(\ref{eq:spatial_augm}), and resetting to background all the label values that we do not wish to segment (i.e., labels for the background structures, which are of no interest to segment). Thus, $T$ has $K' \leq K$ labels (Figure~\ref{fig:generative model}f).

We emphasise that the central contribution of this work lies in the adopted domain randomisation strategy. The values of the hyperparameters controlling the uniform priors (listed in Supplement~2) are tuned using a validation set, and are the object of a sensitivity analysis in Section~\ref{sec:exp 2}.

\subsection{Segmentation network and learning}
\label{sec:learning}

Given the described generative model, a segmentation network is trained by sampling pairs $\{I, T\}$ on the fly. Here we employ a 3D UNet architecture~\citep{ronneberger_u-net_2015} that we used in previous works with synthetic scans~\citep{billot_learning_2020}. Specifically, it consists of five levels, each separated by a batch normalisation layer~\citep{ioffe_batch_2015} along with a max-pooling (contracting path), or upsampling operation (expanding path). All levels comprise two convolution layers with $3\times3\times3$ kernels. Every convolutional layer is associated with an Exponential Linear Unit activation~\citep{clevert_fast_2016}, except for the last one, which uses a softmax. While the first layer counts 24 feature maps, this number is doubled after each max-pooling, and halved after each upsampling. Following the UNet architecture, we use skip connections across the contracting and expanding paths. Note that the network architecture is not a focus of this work: while we employ a UNet~\citep{ronneberger_u-net_2015} (the most widespread network for medical images), it could in principle be replaced with any other segmentation architecture.

We use the soft Dice loss for training~\citep{milletari_v-net_2016}:
\begin{equation}
\label{eq:soft dice}
    \text{Loss}(Y,T) = 1 - \sum_{k=1}^{K'} \frac{2 \times \sum_{x,y,z} Y_k(x,y,z) T_k(x,y,z)}{\sum_{x,y,z} Y_k(x,y,z)^2 \! + \! T_k(x,y,z)^2} \; ,
\end{equation}
where $Y_k$ is the soft prediction for label $k \in \{1, ... K'\}$, and $T_k$ is its associated ground truth in one-hot encoding. We use the Adam optimiser~\citep{kingma_adam_2017} for 300,000 steps with a learning rate of $10^{-4}$, and a batch size of 1. The network is trained twice, and the weights are saved every 10,000 steps. The retained model is then selected relatively to a validation set. In practice, the generative model and the segmentation network are concatenated within a single model, which is entirely implemented on the GPU in Keras~\citep{chollet_keras_2015} with a Tensorflow backend~\citep{abadi_tensorflow_2016}. In total, training takes around seven days on a Nvidia Quadro RTX 6000 GPU.

\subsection{Inference}
\label{sec:inference}

At test time, the input is resampled to $r_{HR}$ with trilinear interpolation (such that the output of the CNN is at HR), and its intensities are rescaled between 0 and 1 with min-max normalisation (using the $1^{st}$ and $99^{th}$ percentiles). Preprocessed scans are then fed to the network to obtain soft predictions maps for each label. In practice, we also perform test-time augmentation~\citep{moshkov_test_2020}, which slightly improved results on the validation set. Specifically, we segment two versions of each test scan: the original one, and a right-left flipped version of it. The soft predictions of the flipped input are then flipped back to native space (while ensuring that right-left labels end up on the correct side), and averaged with the predictions of the original scan. Once test-time augmentation has been performed, final segmentations are obtained by keeping the biggest connected component for each label. On average, inference takes ten seconds on a Nvidia TitanXP GPU (12GB), including preprocessing, prediction, and postprocessing.


\section{General Experimental Setup}

\subsection{Brain scans and ground truths}
\label{sec:datasets}

Our experiments employ eight datasets comprising 5,000 scans of six different modalities and ten resolutions. The splits between training, validation, and testing are given in Table~\ref{tab:datasets}.

\vspace{1mm}
\noindent\textbf{T1-39:} 39 T1-weighted (T1) scans with manual labels for 30 structures~\citep{fischl_whole_2002}. They were acquired with an MP-RAGE sequence at \SI{1}{\milli\meter} isotropic resolution.

\vspace{1mm}
\noindent\textbf{HCP:} 500 T1 scans of young subjects from the Human Connectome Project~\citep{van_essen_human_2012}, acquired at \SI{0.7}{\milli\meter} resolution, and that we resample at \SI{1}{\milli\meter} isotropic resolution.

\vspace{1mm}
\noindent\textbf{ADNI:} 1,500 T1 scans from the Alzheimer's Disease Neuroimaging Initiative (ADNI\footnote{
The ADNI was launched in 2003 by the National Institute on Ageing, the National Institute of Biomedical Imaging and Bioengineering, the Food and Drug Administration, pharmaceutical companies and non-profit organisations, as a 5-year public-private partnership. The goal of ADNI is to test if MRI, PET, other biological markers, and clinical and neuropsychological assessment can analyse the progression of MCI and early AD, develop new treatments, monitor their effectiveness, and decrease the time and cost of clinical trials.}).
All scans are acquired at \SI{1}{\milli\meter} isotropic resolution from a wide array of scanners and protocols. In contrast to HCP, this dataset comprises ageing subjects, some diagnosed with mild cognitive impairment (MCI) or Alzheimer's Disease (AD). As such, many subjects present strong atrophy patterns and white matter lesions. 

\vspace{1mm}
\noindent\textbf{T1mix:} 1,000 T1 scans at \SI{1}{\milli\meter} isotropic resolution from seven datasets: ABIDE~\citep{di_martino_autism_2014}, ADHD200~\citep{the_adhd-200_consortium_adhd-200_2012}, GSP~\citep{holmes_brain_2015}, HABS~\citep{dagley_harvard_2017}, MCIC~\citep{gollub_mcic_2013}, OASIS~\citep{marcus_open_2007}, and PPMI~\citep{marek_parkinson_2011}. We use this heterogeneous dataset to assess robustness against intra-modality contrast variations due to different acquisition protocols.

\vspace{1mm}
\noindent\textbf{FSM:} 18 subjects with T1 and two other MRI contrasts: T2-weighted (T2) and a sequence used for deep brain stimulation (DBS)~\citep{iglesias_probabilistic_2018}. All scans are at \SI{1}{\milli\meter} resolution.

\vspace{1mm}
\noindent\textbf{MSp:} 8 subjects with T1 and proton density (PD) acquisitions at \SI{1}{\milli\meter} isotropic resolution~\citep{fischl_sequence-independent_2004}. These scans were skull stripped prior to availability, and are manually delineated for the same labels as T1-39.

\vspace{1mm}
\noindent\textbf{FLAIR:} 2,393 fluid-attenuated inversion recovery (FLAIR) scans at $1\times 1\times \SI{5}{\milli\meter}$ axial resolution. These subjects are from another subset of the ADNI database, and hence also present morphological patterns related to ageing and AD. This dataset enables assessment on scans that are representative of clinical acquisitions with real-life slice selection profiles (as opposed to simulated LR, see below). Matching \SI{1}{\milli\meter} T1 scans are also available, but they are not used for testing.

\vspace{1mm}
\noindent\textbf{CT:} 6 computed tomography (CT) scans at $1\times1\times\SI{3}{\milli\meter}$ axial resolution~\citep{west_comparison_1997}, with the aim of assessing SynthSeg on 
imaging modalities other than MRI. As for the FLAIR dataset, matching \SI{1}{\milli\meter} T1 scans are also available.

\begin{table}[t]
\centering
\setlength\tabcolsep{3.3pt}
    \caption{Summary of the employed brain datasets. The 10 test resolutions are 1mm$^3$ and 3/5/7mm in either axial, coronal, or sagittal direction. SynthSeg is trained solely on label maps (no intensity images).}
    \small
    \begin{tabular}{c c c c}
        \toprule
        Dataset & Subjects & Modality & Resolution \\
        \midrule
        \multicolumn{4}{c}{Training} \\
        \midrule
        \multirow{2}{*}{T1-39} & \multirow{2}{*}{20} & SynthSeg: Label maps & \multirow{2}{*}{\SI{1}{\milli\meter} isotropic} \\
        && Baselines: T1 \\
        \multirow{2}{*}{HCP} & \multirow{2}{*}{500} & SynthSeg: Label maps & \multirow{2}{*}{\SI{1}{\milli\meter} isotropic} \\
        && Baselines: T1 \\
        \multirow{2}{*}{ADNI} & \multirow{2}{*}{500} & SynthSeg: Label maps & \multirow{2}{*}{\SI{1}{\milli\meter} isotropic} \\
        && Baselines: T1 \\
        \midrule
        \multicolumn{4}{c}{Validation} \\
        \midrule
        T1-39 & 4 & T1 & all tested resolutions\\
        FSM & 3 & T2, DBS & all tested resolutions \\
        \midrule
        \multicolumn{4}{c}{Testing} \\
        \midrule
        T1-39 & 15 & T1 & all tested resolutions \\
        ADNI & 1,000 & T1 & \SI{1}{\milli\meter} isotropic \\
        T1mix & 1,000 & T1 & \SI{1}{\milli\meter} isotropic \\
        FSM & 15 & T1, T2, DBS & all tested resolutions \\
        MSp & 8 & T1, PD & all tested resolutions \\
        FLAIR & 2,393 & FLAIR & \SI{5}{\milli\meter} axial \\
        CT & 6 & CT & \SI{3}{\milli\meter} axial \\
        \bottomrule
    \label{tab:datasets}
    \end{tabular}
\end{table}

\vspace{1mm}
In order to evaluate SynthSeg on more resolutions, we artificially downsample all modalities from the T1-39, FSM, and MSp datasets (all at \SI{1}{\milli\meter} isotropic resolution) to nine different LR: 3, 5 and \SI{7}{\milli\meter} spacing in axial, coronal, and sagittal directions. These simulations do not use real-life slice selection profiles, but are nonetheless very informative since they enable to study the segmentation accuracy as a function of resolution. 

Except for T1-39 and MSp, which are available with manual labels, segmentation ground truths are obtained by running FreeSurfer~\citep{fischl_freesurfer_2012} on the T1 scans of each dataset, and undergo a thorough visual quality control to ensure anatomical correctness. FreeSurfer has been shown to be very robust across numerous independent T1 datasets and yields Dice scores in the range of 0.85-0.88~\citep{fischl_whole_2002, tae_validation_2008}. Therefore, its use as silver standard enables reliable assessment of Dice below 0.85; any scores above that level are considered equally good. Crucially, using FreeSurfer segmentations enables us to evaluate SynthSeg on vast amounts of scans with very diverse contrasts and resolutions, which would have been infeasible with manual tracings only.

\subsection{Training segmentations and population robustness}
\label{sec:label maps}

As indicated in Table~\ref{tab:datasets}, the training set for SynthSeg comprises 20 label maps from T1-39, 500 from HCP, and 500 from ADNI. Mixing these label maps considerably increases the morphological variety of the synthetic scans (far beyond the capacity of the proposed spatial augmentation alone), and thus enlarges the robustness of SynthSeg to a wide range of populations. We emphasise that using automated label maps for training is possible because synthetic images are by design perfectly aligned with their segmentations. We highlight that the training data does not include any real scan.

Because SynthSeg requires modelling all tissue types in the images, we complement the training segmentations with extra-cerebral labels (Supplement~3) obtained with a Bayesian segmentation approach~\citep{puonti_accurate_2020}. Note that these new labels are dropped with 50\% chances during generation, to make SynthSeg compatible with skull stripped images. Moreover, we randomly ``paste'' lesion labels from FreeSurfer with 50\% probability, to build robustness against white matter lesions (Supplement~4). Finally, we further increase the variability of the training data by randomly left/right flipping segmentations and cropping them to $160^3$ volumes.

\subsection{Competing methods}
\label{sec:competing methods}

\noindent We compare SynthSeg against five other approaches:

\vspace{1mm}
\noindent\textbf{T1 baseline~\citep{zhang_generalizing_2020}:} A supervised network trained on real T1 scans (Table~\ref{tab:datasets}). This baseline seeks to assess the performance of supervised CNNs on their source domain, as well as their generalisation to intra-modality (T1) contrast variations. For comparison purposes, we use the same UNet architecture and augmentation scheme (i.e., spatial deformation, intensity augmentation, bias field corruption) as for SynthSeg.

\vspace{1mm}
\noindent\textbf{nnUNet~\citep{isensee_nnu-net_2021}\footnote{https://github.com/MIC-DKFZ/nnUNet}}: A state-of-the-art supervised approach, very similar to the T1 baseline, except that the architecture, augmentation, pre- and postprocessing are automated with respect to the (real) T1 input data.

\vspace{1mm}
\noindent\textbf{Test-time adaptation (TTA)~\citep{karani_test-time_2021}\footnote{https://github.com/neerakara/test-time-adaptable-neural-networks-for-domain-generalization}}: A state-of-the-art domain adaptation method relying on fine-tuning. Briefly, this strategy uses three CNN modules: an image normaliser (five convolutional layers), a segmentation UNet, and a denoising auto-encoder (DAE). At first, the normaliser and the UNet are jointly trained on supervised data of a source domain, while the DAE is trained separately to correct erroneous segmentations. At test time, the UNet and DAE are frozen, and the normaliser is fine-tuned on scans from different target domains by using the denoised predictions of the UNet as ground truth.

\vspace{1mm}
\noindent\textbf{SIFA~\citep{chen_synergistic_2019}\footnote{https://github.com/cchen-cc/SIFA}}: A state-of-the-art unsupervised domain adaptation strategy, where image-to-image translation and segmentation modules are jointly trained (with shared layers). SIFA seeks to align each target domain to the source data in both feature and image spaces for improved adaptation.

\vspace{1mm}
\noindent\textbf{SAMSEG~\citep{puonti_fast_2016}:} A state-of-the-art Bayesian segmentation framework with unsupervised likelihood. As such, SAMSEG is contrast-adaptive, and can segment at any resolution, albeit not accounting for PV effects. SAMSEG does not need to be trained as it solves an optimisation problem for test each scan. Here we use the version distributed with FreeSurfer 7.0, which runs in approximately 15 minutes.

\vspace{1mm}
The predictions of all methods are postprocessed as in Section~\ref{sec:inference}, except for nnUNet, which uses its own postprocessing. We use the default implementation for all competing methods, except for a few minor points that are listed in Supplement~5. All learning-based methods are trained twice, and models are chosen relatively to the validation set. Segmentations are assessed by computing (hard) Dice scores and the $95^{th}$ percentile of the surface distance (SD95, in millimetres).


\begin{table*}[th!]
\centering
\setlength \tabcolsep{4.6pt}
\caption{Mean Dice scores and $95^{th}$ percentile surface distances (SD95) obtained by all methods for every dataset. The best score for each dataset is in bold, and marked with a star if significantly better than all other methods at a 5\% level (two-sided Bonferroni-corrected non-parametric Wilcoxon signed-rank test). Supervised methods cannot segment non-T1 modalities, and domain adaptation strategies are not tested on the source domain.}
\small
\begin{tabular}{l l c c c c c c c c c c}
\toprule 
& & T1-39 & ADNI & T1mix & FSM-T1 & MSp-T1 & FSM-T2 & FSM-DBS & MSp-PD & FLAIR & CT\\
\midrule 
\multirow{2}{*}{T1 baseline} & Dice & 0.91 & 0.83 & 0.86 & 0.84 & 0.82 & - & - & - & - & -\\ 
& SD95 & 1.31 & 2.63 & 2.14 & 2.09 & 3.55 & - & - & - & - & -\\ 
\midrule 
\multirow{2}{*}{nnUNet~\citep{isensee_nnu-net_2021}} & Dice & 0.91 & 0.82 & 0.84 & 0.84 & 0.81 & - & - & - & - & -\\ 
& SD95 & 1.31 & 2.8 & 2.32 & 2.11 & 3.71 & - & - & - & - & -\\ 
\midrule 
\multirow{2}{*}{TTA~\citep{karani_test-time_2021}} & Dice & - & 0.83 & 0.87 & 0.87 & 0.85 & 0.82 & 0.71 & 0.8 & 0.71 & 0.46\\ 
& SD95 & - & 2.26 & 1.73 & 1.72 & 2.14 & 2.35 & 4.48 & 3.71 & 3.95 & 19.43\\ 
\midrule 
\multirow{2}{*}{SIFA~\citep{chen_synergistic_2019}} & Dice & - & 0.8 & 0.82 & 0.84 & 0.84 & 0.82 & 0.82 & 0.74 & 0.73 & 0.62\\ 
& SD95 & - & 3.03 & 2.24 & 2.21 & 2.57 & 2.32 & 2.09 & 4.41 & 3.30 & 4.51\\ 
\midrule 
\multirow{2}{*}{SAMSEG~\citep{puonti_fast_2016}} & Dice & 0.85 & 0.81 & 0.86 & 0.86 & 0.83 & 0.82 & 0.81 & 0.81 & 0.64 & 0.71\\ 
& SD95 & 1.85 & 3.09 & 1.77 & 1.81 & 2.47 & 2.21 & 2.34 & 2.99 & 3.67 & 3.36\\ 
\midrule 
\multirow{2}{*}{SynthSeg (ours)} & Dice & 0.88 & \textbf{0.84} & 0.87 & \textbf{0.88} & \textbf{0.86*} & \textbf{0.86*} & \textbf{0.86*} & \textbf{0.84*} & \textbf{0.78*} & \textbf{0.76*}\\ 
& SD95 & 1.5 & \textbf{2.18*} & \textbf{1.69*} & \textbf{1.59*} & \textbf{1.89*} & \textbf{1.83*} & \textbf{1.81*} & \textbf{2.06*} & \textbf{2.35*} & \textbf{3.29*}\\ 
\bottomrule 
\label{tab:exp1_dice}
\end{tabular}
\end{table*}

\section{Experiments and Results}

Here we present four experiments that evaluates the accuracy and generalisation of SynthSeg. First, we compare it against all competing methods on every dataset. Then, we conduct an ablation study on the proposed method. The third experiment validates SynthSeg in a proof-of-concept neuroimaging group study. Finally, we demonstrate the generalisability of our method by extending it to cardiac MRI and CT.

\subsection{Robustness against contrast and resolution}
\label{sec:exp 1}

\begin{figure}[t]
    \centering
    \includegraphics[width=\columnwidth]{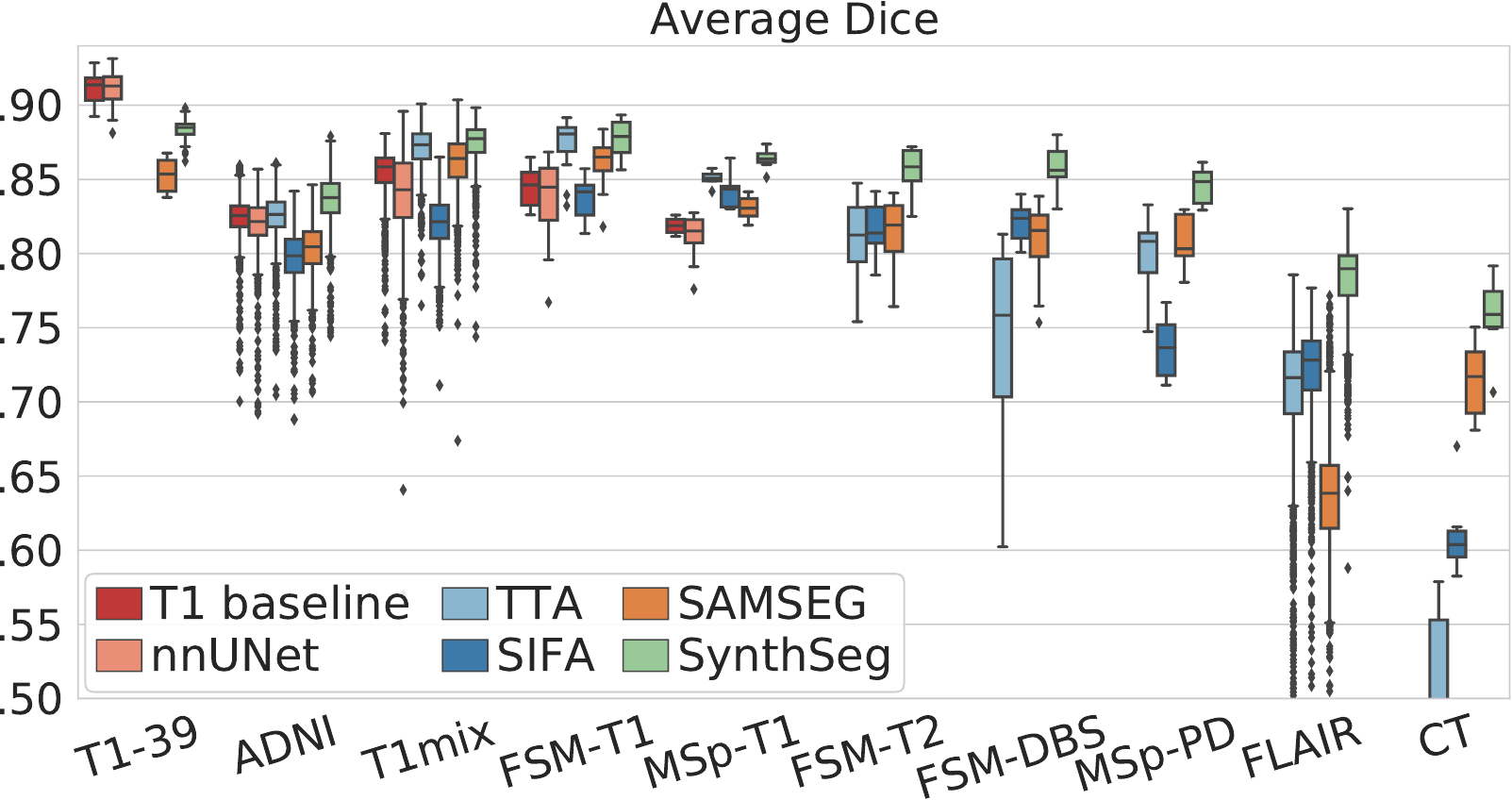}
    \caption{Box plots showing Dice scores obtained by all methods for every dataset. For each box, the central mark is the median; edges are the first and third quartiles; and outliers are marked with $\blacklozenge$.}
    \label{fig:exp1_dice}
\end{figure}

In this experiment, we assess the generalisation ability of SynthSeg by comparing it against all competing methods for every dataset at native resolution (Figure~\ref{fig:exp1_dice}, Table~\ref{tab:exp1_dice}).

\begin{figure}[t]
    \centering
    \includegraphics[width=\columnwidth]{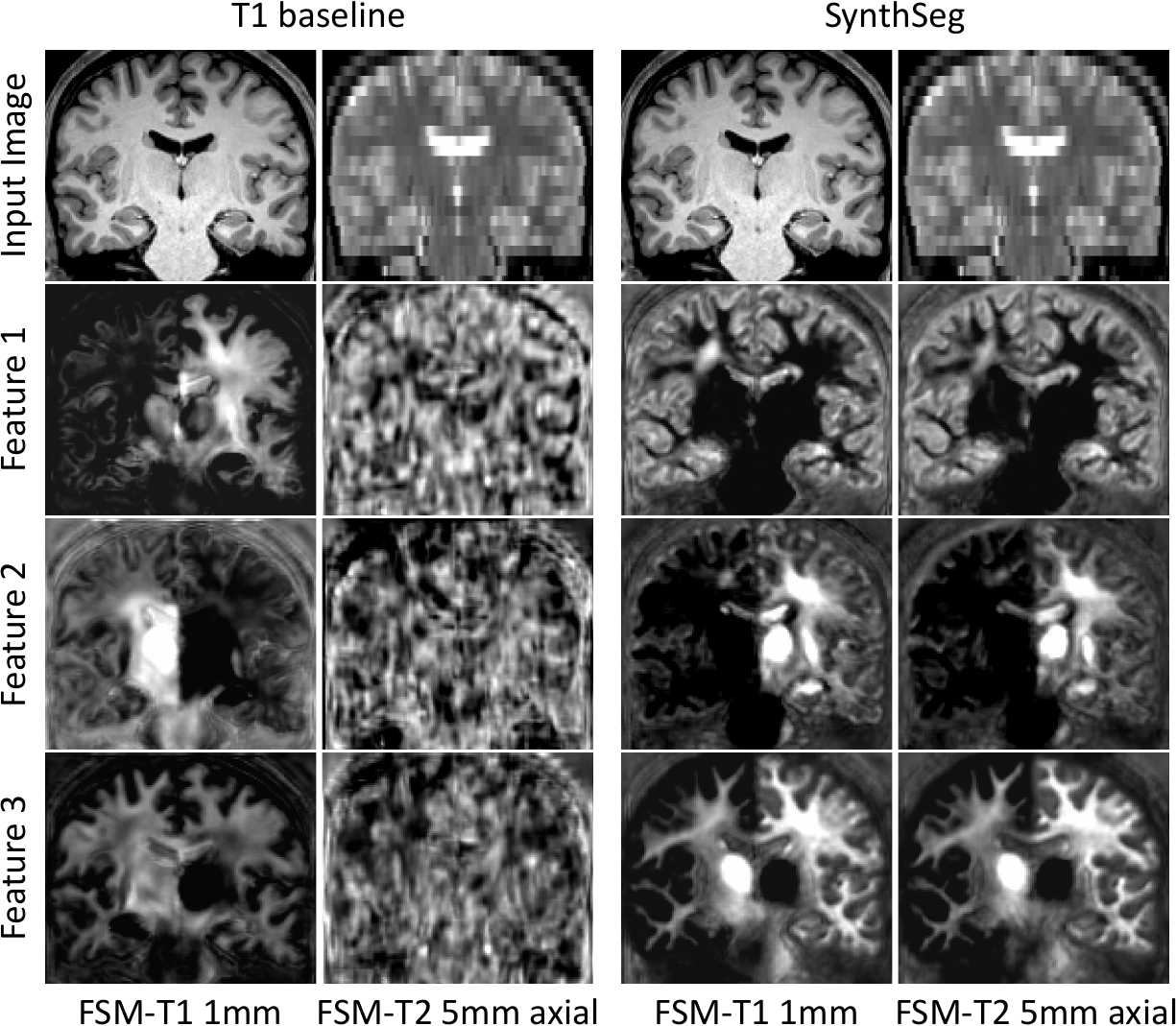}
    \caption{Representative features of the last layer of the network for two scans of different contrast and resolution for the same subject. While the T1 baseline only produces noise outside its training domain, SynthSeg learns a consistent representation across contrasts and resolutions.}
    \label{fig:feat}
\end{figure}

\begin{figure*}[t]
    \centering
    \includegraphics[width=0.97\textwidth]{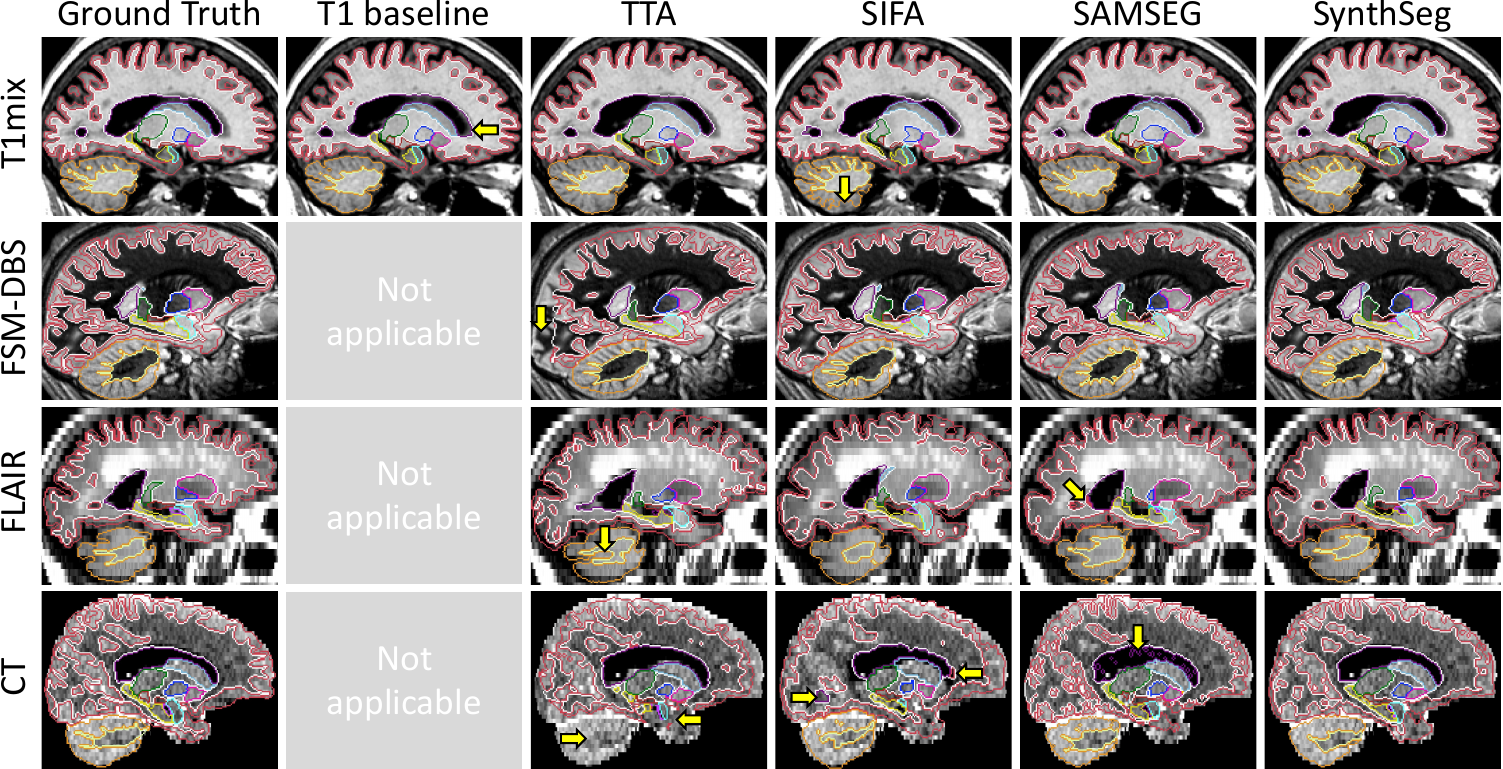}
    \caption{Sample segmentations from the first experiment. Major segmentation mistakes are indicated with yellow arrows. SynthSeg produces very accurate segmentations for all contrasts and resolutions. The T1 baseline makes small errors outside its training domain and cannot be applied to other modalities. While the TTA approach yields very good segmentations for T1mix, its results degrade for larger domain gaps, where it is outperformed by SIFA. Finally, SAMSEG yields coherent results for scans at 1mm resolution, but is heavily affected by PV effects at low resolution.}
    \label{fig:exp1_segs}
\end{figure*}

Remarkably, despite SynthSeg has never been exposed to a real image during training, it reaches almost the same level of accuracy as supervised networks (T1 baseline and nnUNet) on their training domain (0.88 against 0.91 Dice scores on T1-39). Moreover, SynthSeg generalises better than supervised networks against intra-modality contrast variations, both in terms of mean (average difference of 2.5 Dice points with the T1 baseline for T1 datasets other than T1-39) and robustness (much higher lower-quartiles for SynthSeg). Crucially, the employed DR strategy yields very good generalisation, as SynthSeg sustains a remarkable accuracy across all tested contrasts and resolutions, which is infeasible with supervised networks alone. Indeed, SynthSeg outputs high-quality segmentations for all domains, even for FLAIR and CT scans at LR (Figure~\ref{fig:exp1_segs}). Quantitatively, SynthSeg produces the best scores for all nine target domains, six of which with statistical significance for Dice and nine for SD95 (Table~\ref{tab:exp1_dice}). This flexibility is exemplified in Figure~\ref{fig:feat}, where SynthSeg produces features that are almost identical for a 1mm T1 and a 5mm axial T2 of the same subject, the latter being effectively super-resolved to \SI{1}{\milli\meter}.

\begin{figure}[t]
    \centering
    \includegraphics[width=\columnwidth]{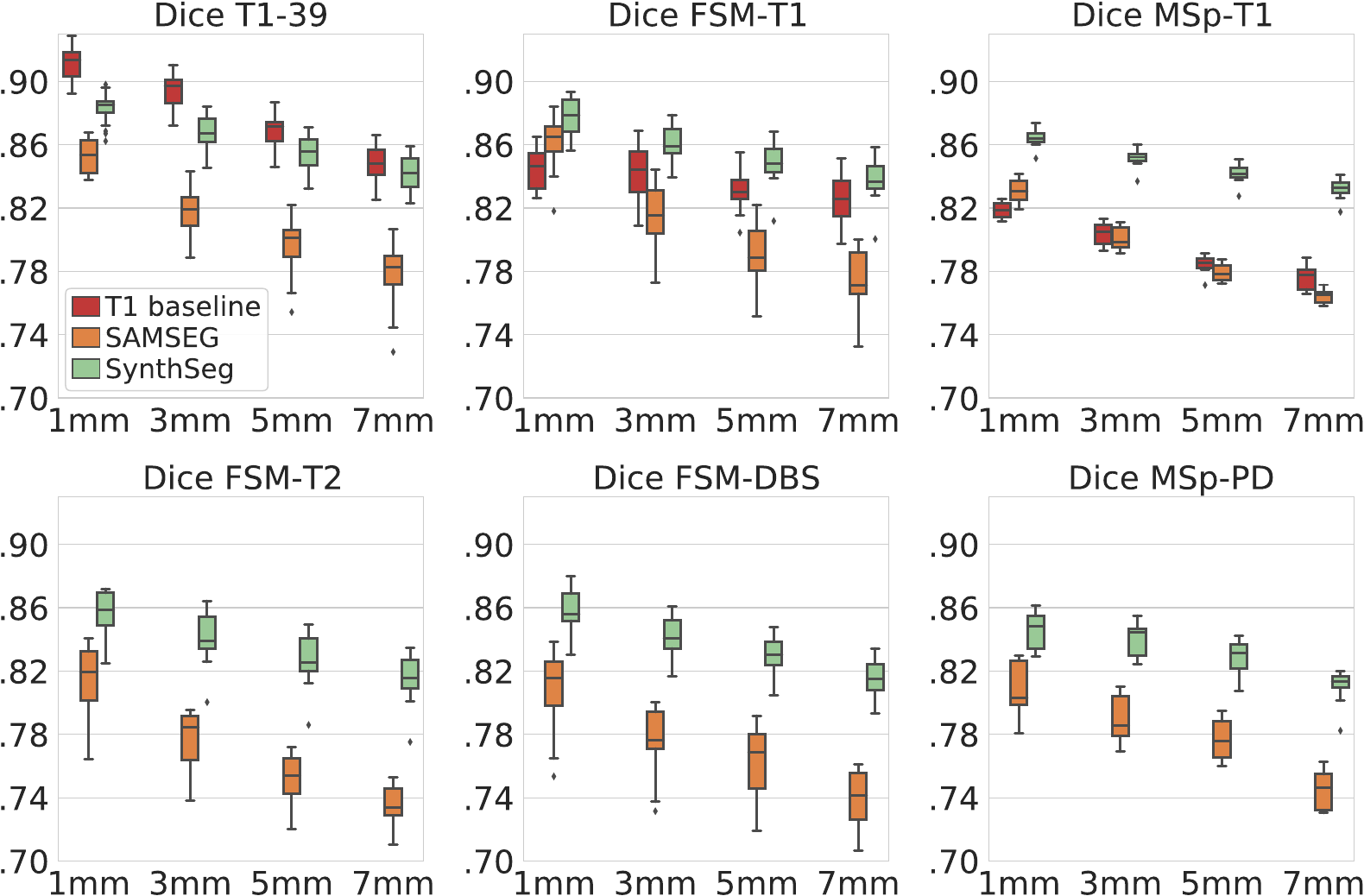}
    \caption{Dice scores for data downsampled at 3, 5, or 7mm in either axial, coronal, or sagittal direction (results are averaged across directions).}
    \label{fig:exp1_resolutions}
\end{figure}

\begin{figure*}[th]
    \centering
    \includegraphics[width=\textwidth]{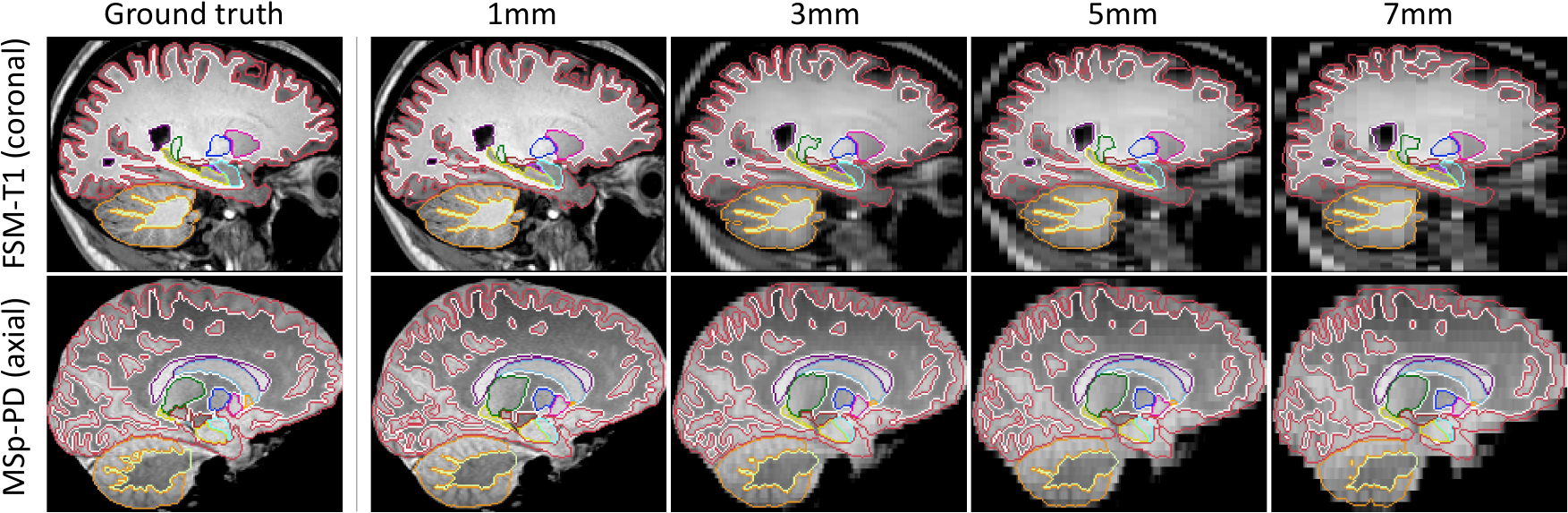}
    \caption{Examples of segmentations obtained by SynthSeg for two scans artificially downsampled at decreasing LR. SynthSeg presents an impressive generalisation ability to all resolutions, despite heavy PV effects and important loss of information at LR. However, we observe a slight decrease in accuracy for thin and convoluted structures such as the cerebral cortex (red) or the white cerebellar matter (dark yellow).}
    \label{fig:ex_res}
\end{figure*}

Although the tested domain adaptation approaches (TTA, SIFA) considerably increase the generalisation of supervised networks, they are still outperformed by SynthSeg for all contrasts and resolutions. This is a remarkable result since, as opposed to domain adaptation strategies, SynthSeg does not require any retraining. We note that fine-tuning the TTA framework makes it more robust than supervised methods for intra-modality applications (noticeably higher lower-quartiles), but its results can substantially fluctuate for larger domain gaps (e.g., on FSM-DBS, FLAIR, and CT). This is partly corrected by SIFA (improvement of 14.92mm in SD95 for CT), which is better suited for larger domain gaps~\citep{karani_test-time_2021}, albeit some abrupt variations (e.g., MSp-PD). In comparison, SAMSEG yields much more constant results across MR contrasts at \SI{1}{\milli\meter} resolution (average Dice score of 0.83). However, because it does not model PV, its accuracy greatly declines at low resolution: Dice scores decrease to 0.71 on the \SI{3}{\milli\meter} CT dataset (5 points below SynthSeg), and to 0.64 on the \SI{5}{\milli\meter} FLAIR dataset (14 points below SynthSeg).

To further validate the flexibility of the proposed approach to different resolutions, we test SynthSeg on all artificially downsampled data (Table~\ref{tab:datasets}), and we compare it against T1 baselines retrained at each resolution, as well as SAMSEG. The results show that SynthSeg maintains a very good accuracy for all tested resolutions (Figure~\ref{fig:exp1_resolutions}). Despite the considerable loss of information at LR and heavy PV effects, SynthSeg only loses 3.8 Dice points between \SI{1}{\milli\meter} and \SI{7}{\milli\meter} slice spacing on average, mainly due to thin structures like the cortex (Figure~\ref{fig:ex_res}). Meanwhile, SAMSEG is strongly affected by PV, and loses 7.6 Dice points across the same range. As before, the T1 baselines obtain remarkable results on scans similar to their training data, but generalise poorly to unseen domains (i.e., FSM-T1 and MSp-T1), where SynthSeg is clearly superior. Moreover, the gap between them on the training data progressively narrows with decreasing resolution, until it almost vanishes at \SI{7}{\milli\meter}, thus making SynthSeg particularly useful for LR scans.

\subsection{Ablations on DR and training label maps}
\label{sec:exp 2}

\begin{figure}[t]
    \centering
    \includegraphics[width=\columnwidth]{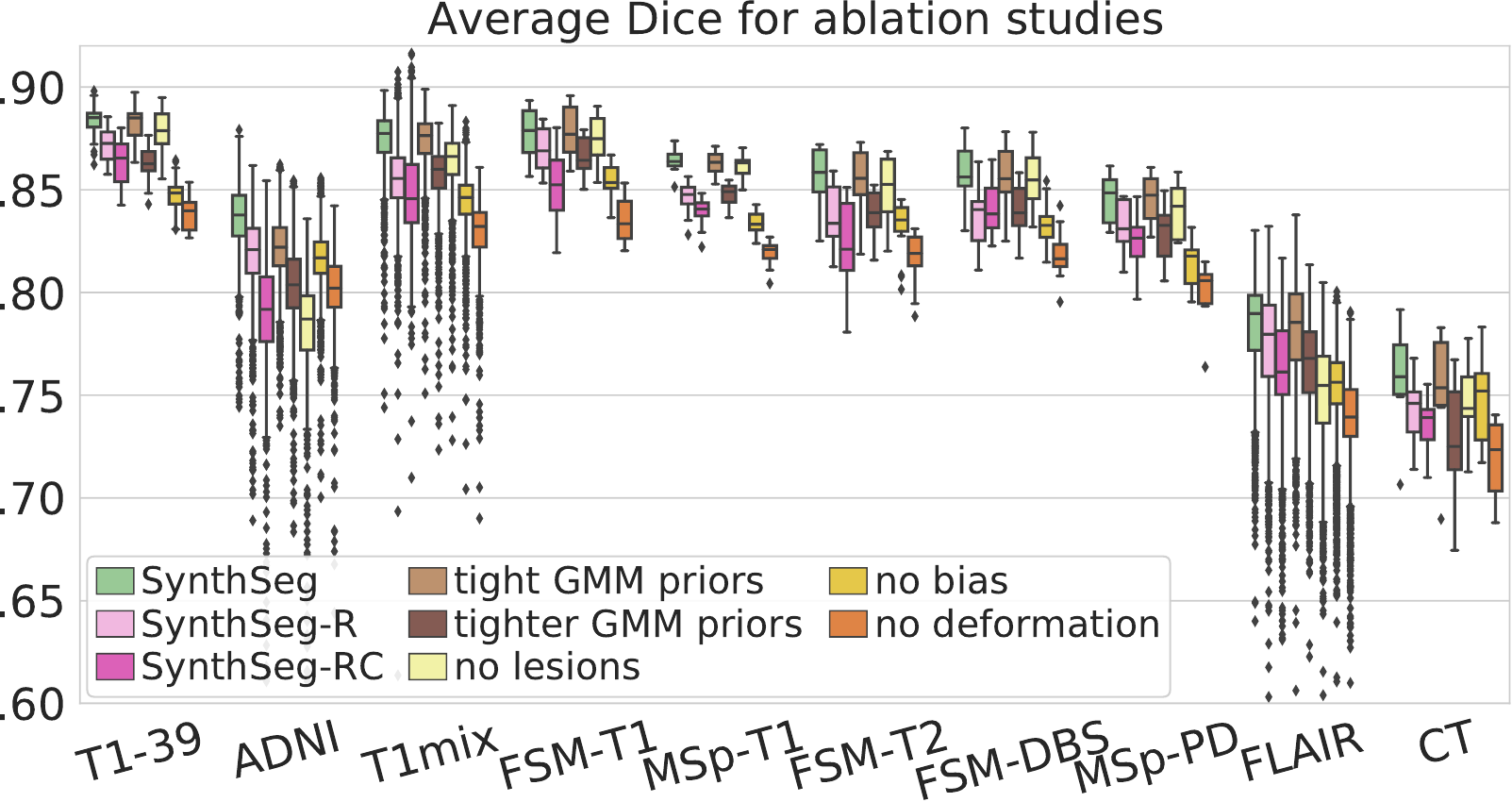}
    \caption{Mean Dice scores obtained for SynthSeg and ablated variants.}
    \label{fig:ablation}
\end{figure}

\begin{figure}[t]
    \centering
    \includegraphics[width=\columnwidth]{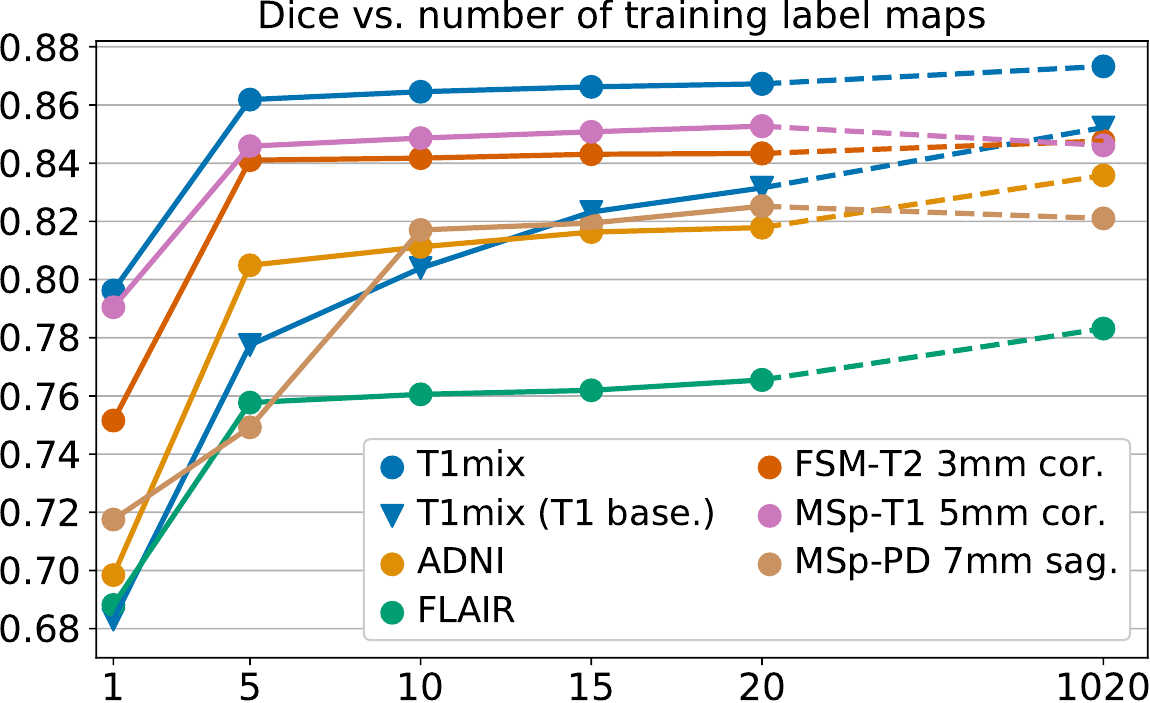}
    \caption{Dice vs. number of training label maps for SynthSeg (circles) on representative datasets. The last points are obtained by training on all available labels maps (20 manual plus 1,000 automated). We also report scores obtained on T1mix by the T1 baseline (triangles).}
    \label{fig:exp3_n_atlas}
\end{figure}

\begin{figure}[t]
    \centering
    \includegraphics[width=\columnwidth]{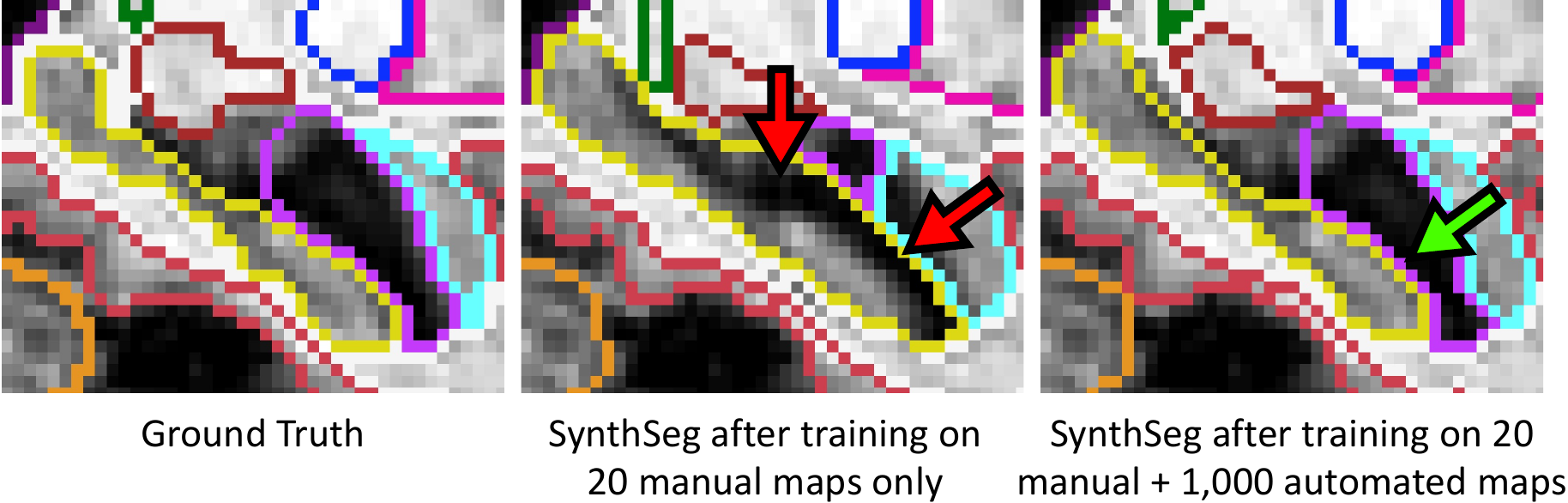}
    \caption{Close-up on the hippocampus for an ADNI testing subject with atrophy patterns that are not present in the manual training segmentations. Hence, training SynthSeg on these manual maps only leads to limited accuracy (red arrows). However, adding a large number of automated maps from different populations to the training set enables us to improve robustness against morphological variability (green arrow).}
    \label{fig:exp3_n_atlas_ex}
\end{figure}

We now validate several aspects of our method, starting with the DR strategy. We first focus on the intensity profiles of the synthetic scans by training four variants: \textit{(i)}~SynthSeg-R, which is resolution-specific; \textit{(ii)}~SynthSeg-RC, which we retrain for every new combination of contrast and resolution by using domain-specific Gaussian priors for the GMM and resolution parameters~\citep{billot_partial_2020,billot_joint_2021}, \textit{(iii)}~a variant using slightly tighter GMM uniform priors ($\mu \in [10, 240]$, $\sigma \in [1, 25]$, instead of $\mu \in [0, 255]$, $\sigma \in [0, 35]$); and \textit{(iv)}~a variant with even tighter priors ($\mu \in [50, 200]$, $\sigma \in [1, 15]$). SynthSeg-R and SynthSeg-RC assess the effect of constraining the synthetic intensity profiles to look more realistic, whereas the two last variants study the sensitivity of the chosen GMM uniform priors. Finally, we train three more networks by ablating the lesion simulation, bias field, and spatial augmentation, respectively.

Figure~\ref{fig:ablation} shows that, crucially, narrowing the distributions of the generated scans in SynthSeg-R and SynthSeg-RC to simulate a specific contrast and/or resolution, leads to a consistent decrease in accuracy: despite retraining them on each target domain, they are on average lower than SynthSeg by 1.4 and 2.6 Dice points respectively. Interestingly, the variant with slightly tighter GMM priors obtains scores almost identical to the reference SynthSeg, whereas further restricting these priors (at the risk of excluding intensities encountered in real scans) leads to poorer performance (2.1 fewer Dice points on average). Finally, the bias and deformation ablations highlight the impact of those two augmentations (loss of 3.7 and 4.3 Dice points, respectively), whereas ablating the lesion simulation mainly affects the ADNI and FLAIR datasets, where the ageing subjects are more likely to present lesions (average loss of 3.9 Dice points).

In a second set of experiments, we evaluate the effect of using different numbers of segmentations during training. Hence, we retrain SynthSeg on increasing numbers of label maps randomly selected from T1-39 ($N \in \{1, 5, 10, 15, 20\}$, see Supplement~6). Moreover, we include the version of SynthSeg trained on all available maps, to quantify the effect of adding automated segmentations from diverse populations. All networks are evaluated on six representative datasets (Figure~\ref{fig:exp3_n_atlas}). The results reveal that using only one training map already attains decent Dice scores (between 0.68 and 0.80 for all datasets). As expected, the accuracy increases when adding more maps, and Dice scores plateau at $N=5$ (except for MSp-PD, which levels off at $N=10$). Interestingly, Figure~\ref{fig:exp3_n_atlas} also shows that SynthSeg requires fewer training examples than the T1 baseline to converge towards its maximum accuracy.

Meanwhile, adding a large amount of training automated maps enables us to improve robustness to morphological variability, especially for the ADNI and FLAIR datasets with ageing and diseased subjects (Dice scores increase by 1.9 and 2.0 points respectively). To confirm this trend, we study the 3\% of ADNI subjects with the largest ventricular volumes (relatively to the intracranial volume, ICV), whose morphology substantially deviates from the 20 manual training maps. For these ADNI cases (30 in total), the average Dice score increases by 4.7 Dice points for the network trained on all label maps compared with the one trained on manual maps only. This result further demonstrates the gain in robustness obtained by adding automated label maps to the training set of SynthSeg (Figure~\ref{fig:exp3_n_atlas_ex}).

\subsection{Alzheimer's Disease volumetric study}
\label{sec:exp 3}

In this experiment, we evaluate SynthSeg in a proof-of-concept volumetric group study, where we assess its ability to detect hippocampal atrophy related to AD~\citep{chupin_fully_2009}. Specifically, we study whether SynthSeg can detect similar atrophy patterns for subjects who have been imaged with different protocols. As such, we run SynthSeg on a separate set of 100 ADNI subjects (50 controls, 50 AD), all with \SI{1}{\milli\meter} isotropic T1 scans as well as FLAIR acquisitions at \SI{5}{\milli\meter} axial resolution. 

We measure atrophy with effect sizes in predicted volumes between controls and diseased populations. Effect sizes are computed with Cohen's~$d$~\citep{cohen_statistical_1988}:
\begin{equation}
d \! = \! \frac{\mu_{C} - \mu_{AD}}{s}, \! \! \! \quad s \! = \! \sqrt{\frac{(n_{C} - 1)s_{C}^{2}  +  (n_{AD} - 1)s_{AD}^{2}}{n_{C}+n_{AD}-2}},
\end{equation}
where $\mu_{C}$, $s_{C}^{2}$ and $\mu_{AD}$, $s_{AD}^{2}$ are the means and variances of the volumes for the two groups, and $n_{C}$ and $n_{AD}$ are their sizes. Hippocampal volumes are computed by summing the corresponding soft predictions, thus accounting for segmentation uncertainties. All measured volumes are corrected for age, gender, and ICV (estimated with FreeSurfer) using a linear model.

\begin{table}[t]
\centering
\setlength\tabcolsep{3pt} 
    \caption{Effect size (Cohen's d) obtained by FreeSurfer (Ground Truth, GT), SAMSEG and SynthSeg for hippocampal volumes between controls and AD patients for different types of scans. }
    \small
    \begin{tabular}{l c c c c c}
        \toprule
         Contrast & Resolution & FreeSurfer (GT) & SAMSEG & SynthSeg  \\
        \midrule
        T1 & \SI{1}{\milli\meter}$^3$ &\multirow{2}{*}{1.38}  & 1.46 & 1.40  \\
        FLAIR & \SI{5}{\milli\meter} axial & & 0.53 & 1.24 \\
        \bottomrule
    \label{tab:cohens_d}
    \end{tabular}
\end{table}

In addition to SynthSeg, we evaluate the performance of SAMSEG, and all Cohen'$d$ are compared to a silver standard obtained by running FreeSurfer on the T1 scans~\citep{fischl_freesurfer_2012}. The results, reported in Table~\ref{tab:cohens_d}, reveal that both methods yield a Cohen's $d$ close to the ground truth for the HR T1 scans. We emphasise that, while segmenting the hippocampus in \SI{1}{\milli\meter} T1 scans is of modest complexity, this task is much more difficult for \SI{5}{\milli\meter} axial FLAIR scans, since the hippocampus only appears in two to three slices, and with heavy PV. As such, the accuracy of SAMSEG greatly degrades on FLAIR scans, where it obtains less than half the expected effect size. In contrast, SynthSeg sustains a high accuracy on the FLAIR scans, producing a Cohen's $d$ much closer to the reference value, which was obtained at HR.

\subsection{Extension to cardiac segmentation}
\label{sec:exp 4}

\begin{figure}[t]
    \centering
    \includegraphics[width=\columnwidth]{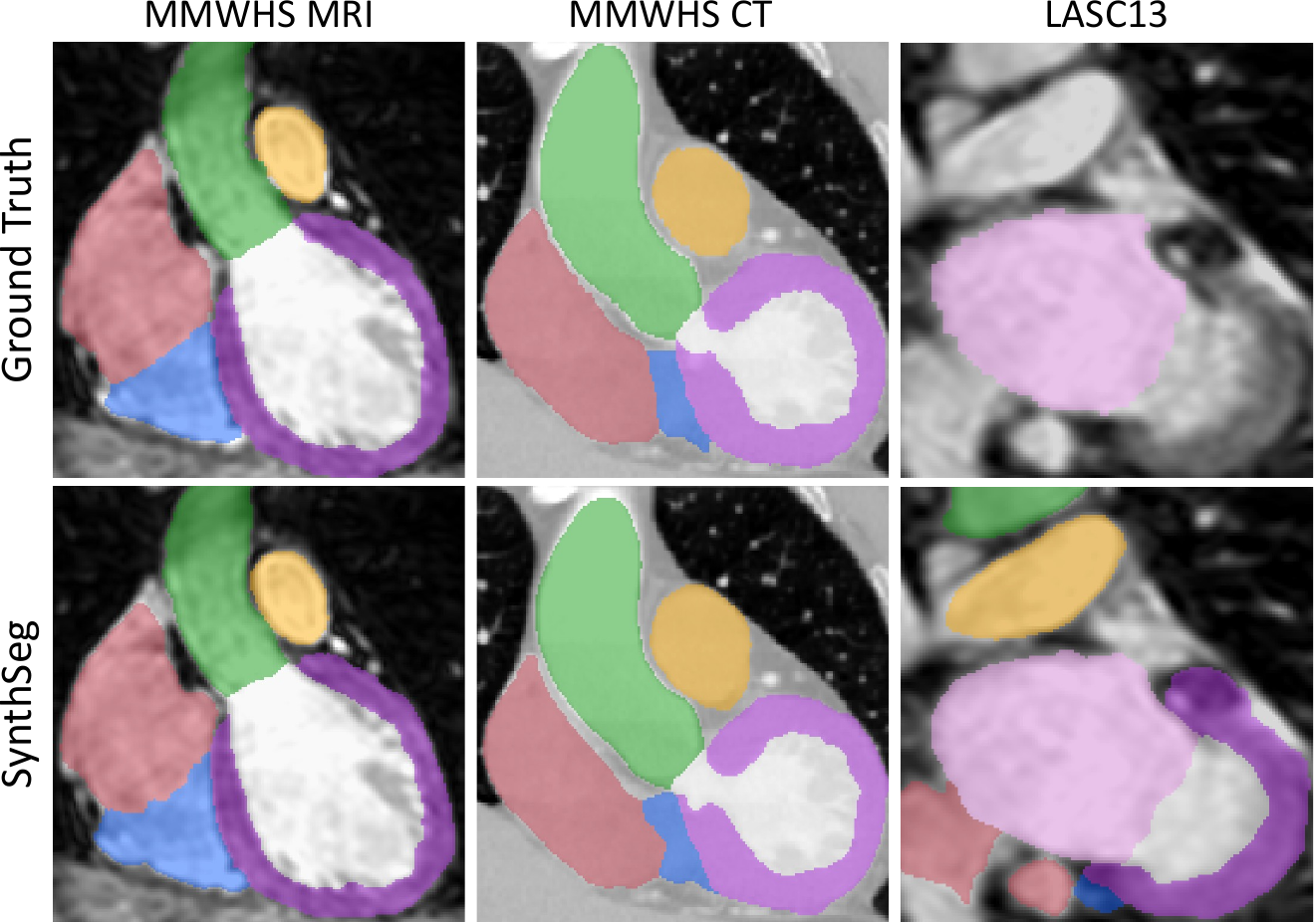}
    \caption{Representative cardiac segmentations obtained by SynthSeg on three datasets, without retraining on any of them, and without using real images during training. LASC13 only has ground truth for LA (pink).}
    \label{fig:cardiac}
\end{figure}

In this last experiment, we demonstrate the generalisability of SynthSeg by applying it to cardiac segmentation. With this purpose, we employ two new datasets: MMWHS~\citep{zhuang_evaluation_2019}, and LASC13~\citep{tobon-gomez_benchmark_2015}. MMWHS includes 20 MRI scans with in-plane resolutions from 0.78 to \SI{1.21}{\milli\meter}, and slice spacings between 0.9 and \SI{1.6}{\milli\meter}. MMWHS also contains 20 CT scans of non-overlapping subjects at high resolution (0.28-\SI{0.58}{\milli\meter} in-plane, 0.45-\SI{0.62}{\milli\meter} slice spacing). All these scans are available with manual labels for seven regions (see Table~\ref{tab:heart}). On the other hand, LASC13 includes 10 MRI heart scans at $1.25 \times 1.25 \times$ \SI{1.37}{\milli\meter} axial resolution, with manual labels for the left atrium only. We form the training set for SynthSeg by randomly drawing 13 label maps from MMWHS MRI. For consistency, these training segmentations are all resampled at a common \SI{1}{\milli\meter} isotropic resolution. Finally, the validation set consists of two more scans from MMWHS, while all the remaining scans are used for testing.

\begin{table}[t]
\centering
\setlength \tabcolsep{4.5pt}
\caption{Dice scores for seven cardiac regions: left atrium (LA), right atrium (RA), left ventricle (LV), right ventricle (RV), myocardium (MYO), ascending aorta (AA), and pulmonary artery (PA). LASC13 only has ground truth for LA. }
\small
\begin{tabular}{l c c c c c c c}
\toprule 
& LA & LV & RA & RV & MYO & AA & PA\\
\midrule 
MMWHS MRI & 0.91 & 0.89 & 0.9 & 0.84 & 0.81 & 0.86 & 0.86\\ 
\midrule 
MMWHS CT & 0.92 & 0.89 & 0.86 & 0.88 & 0.85 & 0.94 & 0.84\\ 
\midrule 
LASC13 & 0.9 & - & - & - & - & - & -\\ 
\bottomrule 
\label{tab:heart}
\end{tabular}
\end{table}

Nevertheless, the training labels maps only model the target regions to segment, whereas SynthSeg requires labels for all the tissues present in the test images. Therefore, we enhance the training segmentations by subdividing all their labels (background and foreground) into finer subregions. This is achieved by clustering the intensities of the associated image with the Expectation Maximisation algorithm~\citep{dempster1977maximum}. First, each foreground label is divided into two regions to model blood pools. Then, the background region is split into a random number of $N$ regions ($N \in [3, 10]$), which aim at representing the surrounding structures with different levels of granularity (Supplement~6). All these label maps are precomputed to alleviate computing resources during training. We also emphasise that all sub-labels are merged back with their initial label for the loss computation during training. The network is trained twice as described in Section~\ref{sec:learning} with hyperparameters values obtained relatively to the validation set (see values in Supplement 8). Inference is then performed as in Section~\ref{sec:inference}, except for the test-time flipping augmentation that is now disabled.

The results are reported in Table~\ref{tab:heart}, and show that SynthSeg segments all seven regions with very high precision (all Dice scores are above 0.8). Moreover, it maintains a very good accuracy across all tested datasets, with mean Dice scores of 0.84 and 0.88 for MMWHS MRI and CT respectively. Interestingly, these scores are similar to the state-of-the-art results in cross-modality cardiac segmentation obtained by \citealt{chen_synergistic_2019} (Dice score of 0.82), despite not being \emph{directly} comparable due to differences in resolution (\SI{2}{\milli\meter} for~\citealt{chen_synergistic_2019}, \SI{1}{\milli\meter} for SynthSeg). Overall, segmenting all datasets at such a level of accuracy (Figure~\ref{fig:cardiac}) is remarkable for SynthSeg, since, as opposed to \citealt{chen_synergistic_2019}, it is not retrained on any of them.

\section{Discussion}

We have proposed a method for segmentation of brain MRI scans that is robust against changes in resolution and contrast (including CT) without retraining or fine-tuning. Our main contribution lies in the adopted domain randomisation strategy, where a segmentation network is trained with synthetic scans of fully randomised contrast and resolution. By producing highly diverse samples that make no attempt at realism, this approach forces the network to learn domain-independent features.

The impact of the DR strategy is demonstrated by the domain-constrained SynthSeg variants, for which training contrast and resolution-specific networks yields poorer performance (Section~\ref{sec:exp 2}). We believe this outcome is likely a combination of two phenomena. First, randomising the generation parameters enables us to mitigate the assumptions made when designing the model (e.g., Gaussian intensity distribution for each region, slice selection profile, etc.). Second, this result is consistent with converging evidence that augmenting the data beyond realism often leads to better generalisation~\citep{tobin_domain_2017,chaitanya_semi-supervised_2019,bengio_deep_2011}.

Additionally, SynthSeg enables to greatly alleviate the labelling labour for training purposes. First, it is only trained once and only requires a single set of anatomical segmentations (no real images), as opposed to supervised methods, which need paired images and labels for every new domain. Second, our results show that SynthSeg typically requires less training examples than supervised CNNs to converge to its maximum performance (Section~\ref{sec:exp 2}). And third, parts of the training dataset can be acquired at almost no cost, by including label maps obtained by segmenting real brain scans with automated methods, and visually checking the results to ensure reasonable quality and anatomical plausibility. We highlight that while automated segmentations are generally not used for training (since they are prone to errors), this is made possible here by the fact that synthetic scans are, by design, perfectly aligned with their ground truths. We also emphasise that using automated segmentations to train SynthSeg is not only possible, but recommended, as the inclusion of such segmentations greatly improves robustness against highly different morphologies caused by anatomical variability (e.g., ageing subjects).

Nevertheless, the employed Gaussian model imposes that the training label maps encompass tracings of all tissues present in the test scans. However, this is not a limitation in practice, since automated labels can be obtained for missing structures by simple intensity clustering. This strategy enabled us to obtain state-of-the-art results for cardiac segmentation, where the original label maps did not describe the complex distribution of finer structures (blood pools in cardiac chambers) and surrounding tissues (vessels, bronchi, bones, etc.). Moreover, our results show that SynthSeg can handle deviations from the Gaussian model within a given structure if they are mild (like the thalamus in brain MRI), or far away from the regions to segment (like the neck in brain MRI).

A limitation of this work is the high proportion of automated label maps used for evaluation. This choice was initially motivated by the wish to evaluate SynthSeg on a wide variety of contrasts and resolutions, which would have been infeasible with manual labels only. Nonetheless, we emphasise that a lot of testing datasets still use manual segmentations (T1-39, and MSp for brain segmentation; MMWHS and LASC13 for the heart experiment), and that the remaining datasets have all undergone thorough visual quality control. Importantly, SynthSeg has shown the same remarkable generalisation ability when evaluated with manual or automated ground truths. Finally, the conclusions of this paper are further reinforced by the indirect evaluation performed in Section~\ref{sec:exp 3}, which demonstrates the accuracy and clinical utility of SynthSeg.

Thanks to its unprecedented generalisation ability, SynthSeg yields direct applications in the analysis of clinical scans, for which no general segmentation routines are available due to their highly variable acquisition procedures (sequence, resolution, hardware). Indeed, current methods deployed in the clinic include running FreeSurfer on companion \SI{1}{\milli\meter} T1 scans and/or using such labels to train a supervised network (possibly with domain adaptation) to segment other sequences. However, these methods preclude the analysis of the majority of clinical datasets, where \SI{1}{\milli\meter} T1 scans are rarely available. Moreover, training neural networks in the clinic is difficult in practice, since it requires corresponding expertise. In contrast, SynthSeg achieves comparable results to supervised CNNs on their training domain (especially at LR), and can be deployed much more easily since it does not need to be retrained.

\section{Conclusion}

In this article, we have presented SynthSeg, a learning strategy for segmentation of brain MRI and CT scans, where robustness against a wide range of contrasts and resolutions is achieved without any retraining or fine-tuning. First, we have demonstrated SynthSeg on 5,000 scans spanning eight datasets, six modalities and 10 resolutions, where it maintains a uniform accuracy and almost attains the performance of supervised CNNs on their training domain. SynthSeg obtains slightly better scores than state-of-the-art domain adaptation methods for small domain gaps, while considerably outperforming them for larger domain shifts. Additionally, the proposed method is consistently more accurate than Bayesian segmentation, while being robust against PV effects and running much faster. SynthSeg can reliably be used in clinical neuroimaging studies, as it precisely detects AD atrophy patterns on HR and LR scans alike. Finally, by obtaining state-of-the-art results in cardiac cross-modality segmentation, we have shown that SynthSeg has the potential to be applied to other medical imaging problems.

While this article focuses on the use of domain randomisation to build robustness against changes in contrast and resolution, future work will seek to further improve the accuracy of the proposed method. As such, we will explore the use of adversarial networks to enhance the quality of the synthetic scans. Then, we plan to investigate the use of CNNs to ``denoise'' output segmentations for improved robustness, and we will examine other architectures to replace the UNet employed in this work. Finally, while the ablation of the lesion simulation in Section~\ref{sec:exp 2} is a first evidence of the robustness of SynthSeg to the presence of lesions, future work will seek to precisely quantify the performance of SynthSeg when exposed to various types of lesions, tumours, and pathologies.

The trained model is distributed with FreeSurfer. Relying on a single model will greatly facilitate the use of SynthSeg by researchers, since it eliminates the need for retraining, and thus the associated requirements in terms of hardware and deep learning expertise. By producing robust and reproducible segmentations of nearly any brain scan, SynthSeg will enable quantitative analyses of huge amounts of existing clinical data, which could greatly improve the characterisation and diagnosis of neurological disorders.

\section*{Acknowledgements}
This research is supported by the European Research Council (ERC Starting Grant 677697, project BUNGEE-TOOLS), the EPSRC-funded UCL Centre for Doctoral Training in Medical Imaging (EP/L016478/1) and the Department of Health’s NIHR-funded Biomedical Centre at UCL Hospitals. 
Further support is provided by Alzheimer's Research UK (ARUKIRG2-019A003), the NIH BRAIN Initiative (RF1MH123195 and  U01MH117023), the National Institute for Biomedical Imaging and Bioengineering (P41EB015896, 1R01EB023281, R01EB006758, R21EB018907, R01EB019956), the National Institute on Aging (1R01AG070988, 1R56AG064027, 5R01A-G008122, R01AG016495, 1R01AG064027), the National Institute of Mental Health the National Institute of Diabetes and Digestive (1R21DK10827701), the National Institute for Neurological Disorders and Stroke (R01NS112161, R01NS0525851, R21NS072652, R01NS070963, R01NS0835-34, 5U01NS086625, 5U24NS10059103, R01NS105820), the Shared Instrumentation Grants (1S10RR023401, 1S10R-R019307, 1S10RR023043), the Lundbeck foundation (R3132019622), the NIH Blueprint for Neuroscience Research (5U01MH093765).

The collection and sharing of the ADNI data was funded by the Alzheimer's Disease Neuroimaging Initiative (National Institutes of Health Grant U01 AG024904) and Department of Defence (W81XWH-12-2-0012). ADNI is funded by the National Institute on Aging, the National Institute of Biomedical Imaging and Bioengineering, and the following: Alzheimer's Association; Alzheimer's Drug Discovery Foundation; BioClinica, Inc.; Biogen Idec Inc.; Bristol-Myers Squibb Company; Eisai Inc.; Elan Pharmaceuticals, Inc.; Eli Lilly and Company; F. Hoffmann-La Roche Ltd and affiliated company Genentech, Inc.; GE Healthcare; Innogenetics, N.V.; IXICO Ltd.; Janssen Alzheimer Immunotherapy Research \& Development, LLC.; Johnson \& Johnson Pharmaceutical Research \& Development LLC.; Medpace, Inc.; Merck \& Co., Inc.; Meso Scale Diagnostics, LLC.; NeuroRx Research; Novartis Pharmaceuticals Corporation; Pfizer Inc.; Piramal Imaging; Servier; Synarc Inc.; and Takeda Pharmaceutical Company. The Canadian Institutes of Health Research is providing funds for ADNI clinical sites in Canada. Private sector contributions are facilitated by the Foundation for the National Institutes of Health. The grantee is the Northern California Institute for Research and Education, and the study is coordinated by the Alzheimer's Disease Cooperative Study at the University of California, San Diego. ADNI is disseminated by the Laboratory for Neuro Imaging at the University of Southern California.

\balance



\onecolumn

{\centering \LARGE SynthSeg: Segmentation of brain MRI scans of any \\ contrast and resolution without retraining \\ Supplementary materials \par}

\vspace{1cm}

{\centering \large Benjamin Billot, Douglas N. Greve, Oula Puonti, Axel Thielscher, Koen Van Leemput, Bruce Fischl, Adrian V. Dalca, and Juan Eugenio Iglesias\par}

\setcounter{page}{1}
\setcounter{figure}{0}
\setcounter{table}{0}
\renewcommand{\thefigure}{S\arabic{figure}}
\renewcommand{\thetable}{S\arabic{table}}

\vspace{1cm}
\section*{Supplement 1: $\:$ Examples of generated synthetic scans}
\begin{figure}[h]
    \centering
    \includegraphics[width=\textwidth]{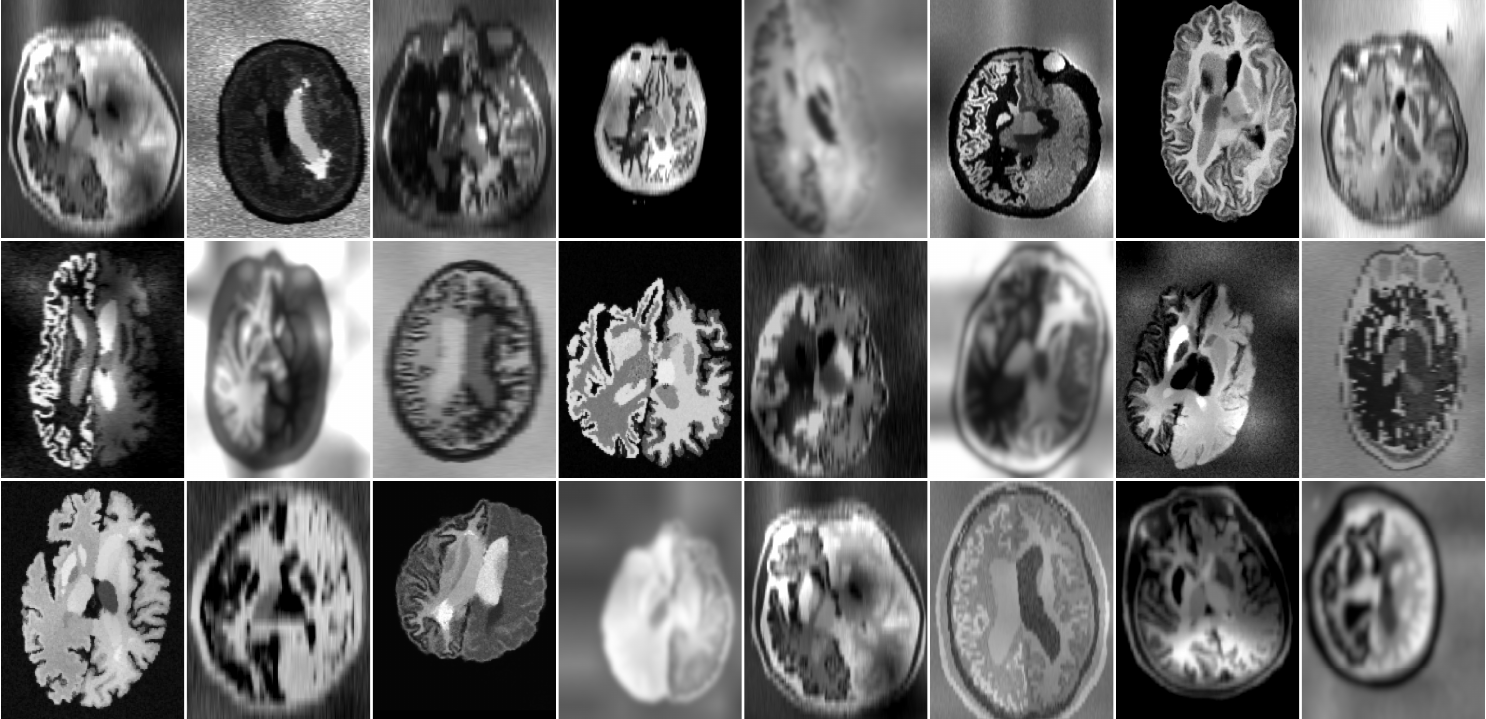}
    \caption{Representative samples from the presented generated model. Synthetic scans present a considerable diversity in terms of contrasts and resolutions, but also in terms of sizes, shapes, bias fields, skull stripping, lesions, and anatomical morphology.
    }
\end{figure}

\vspace{6cm}

\pagebreak

\section*{Supplement 2: $\:$ Values of the generative model hyperparameters}

\vspace{-2cm}

\begin{table}[h]
\centering
\setlength\tabcolsep{4pt} 
    \caption{Values of the hyperparameters controlling the generative model. Intensity parameters assume an input in the [0, 255] interval. Rotations are expressed in degrees, and spatial measures are in millimeters.}
    \medskip
    \small
    \begin{tabular}{lccccccccccccccccccc}
        \toprule
        Hyperparameter & $a_{rot}$ & $b_{rot}$ & $a_{sc}$ & $b_{sc}$ & $a_{sh}$ & $b_{sh}$ & $a_{tr}$ & $b_{tr}$ & $b_{nonlin}$ & $a_\mu$ & $b_\mu$ & $a_\sigma$ & $b_\sigma$ & $b_B$ & $\sigma_\gamma^2$ & $r_{HR}$ & $b_{res}$ & $a_\alpha$ & $b_\alpha$ \\
        \midrule
        Value          &     -20   &    20     &   0.8   &   1.2   &  -0.015  &   0.015  &   -30    &   30     &      4       &   0    &    255  &      0     &     35     &  0.6  &        0.4        &           1         &     9     &    0.95    &     1.05   \\
        \bottomrule
    \end{tabular}
\end{table}

\vspace{1cm}

\section*{Supplement 3: $\:$ List of label values used during training}

\vspace{-2cm}

\begin{table}[h]
\centering
\setlength\tabcolsep{9pt} 
    \caption{List of the labels used for image synthesis, prediction, and evaluation. Note that during generation, we randomly model skull stripping with 50\% probability, by removing all extra-cerebral labels from the training segmentation. In addition, we also model \emph{imperfect} skull stripping, with a further 50\% chances (so 25\% of the total cases), by removing all the extra-cerebral labels except the cerebro-spinal fluid (which surrounds the brain). Different contralateral labels are used for structures marked with~$^{\text{R/L}}$.}
    \medskip
    \small
    \begin{tabular}{l c c c}
        \toprule
        \multirow{2}{*}{Label} & removed for skull & \multirow{2}{*}{predicted} & \multirow{2}{*}{evaluated} \\  
        & stripping simulation & & \\
        \midrule
        Background  & N/A & yes & no \\
        Cerebral white matter$^{\text{R/L}}$ & no & yes & yes \\
        Cerebral cortex$^{\text{R/L}}$  & no & yes & yes  \\
        Lateral ventricle$^{\text{R/L}}$  & no & yes & yes  \\
        Inferior Lateral Ventricle$^{\text{R/L}}$  & no & yes & no  \\
        Cerebellar white matter$^{\text{R/L}}$  & no & yes & yes   \\
        Cerebellar grey matter$^{\text{R/L}}$  & no & yes & yes   \\
        Thalamus$^{\text{R/L}}$  & no & yes & yes   \\
        Caudate$^{\text{R/L}}$  & no & yes & yes  \\
        Putamen$^{\text{R/L}}$  & no & yes & yes   \\
        Pallidum$^{\text{R/L}}$  & no & yes & yes  \\
        Third ventricle   & no & yes & yes  \\
        Fourth ventricle   & no & yes & yes  \\
        Brainstem   & no & yes & yes  \\
        Hippocampus$^{\text{R/L}}$  & no & yes & yes   \\
        Amygdala$^{\text{R/L}}$   & no & yes & yes  \\
        Accumbens area$^{\text{R/L}}$   & no & yes & no  \\
        Ventral DC$^{\text{R/L}}$  & no & yes & no  \\
        Cerebral vessels$^{\text{R/L}}$  & no & no & no   \\
        Choroid plexus$^{\text{R/L}}$   & no & no & no  \\
        White matter lesions$^{\text{R/L}}$  & no & no & no  \\
        Cerebro-spinal Fluid (CSF)  & yes/no & no & no  \\
        Artery  & yes & no & no   \\
        Vein  & yes & no & no   \\
        Eyes   & yes & no & no  \\
        Optic nerve  & yes & no & no  \\
        Optic chiasm   & yes & no & no  \\
        Soft tissues   & yes & no & no  \\
        Rectus muscles  & yes & no & no   \\
        Mucosa   & yes & no & no  \\
        Skin  & yes & no & no   \\
        Cortical bone   & yes & no & no  \\
        Cancellous bone  & yes & no & no   \\
        \bottomrule
    \end{tabular}
\end{table}

\vspace{1cm}

\pagebreak

\section*{Supplement 4: $\:$ Versions of the training label maps}

\begin{figure}[h]
    \centering
    \includegraphics[width=0.8\textwidth]{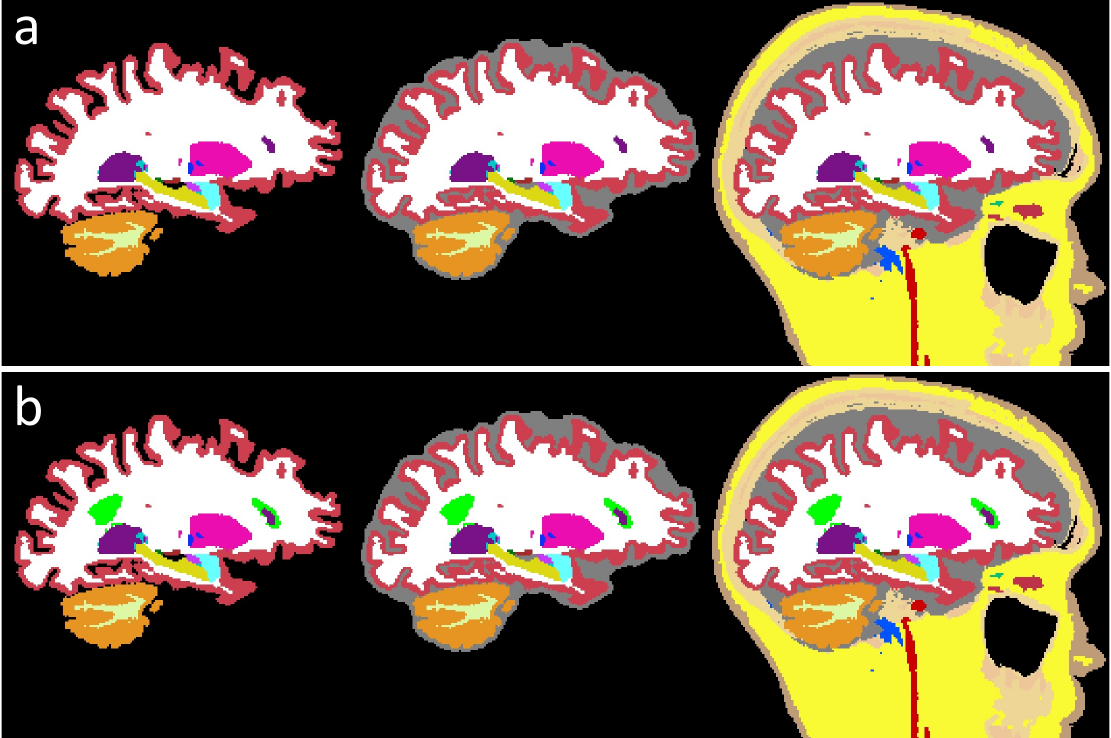}
    \caption{Example of all versions of the label maps used during training. Each map is available (a)~with, or (b)~without lesion labels (bright green), and at different levels of skull stripping: perfect (left), imperfect (middle), or no skull stripping (right). Using these different versions of training label maps enables us to build robustness to white matter lesions and to (possibly imperfect) skull stripping. The lesion labels are obtained with FreeSurfer~\citep{fischl_freesurfer_2012} and directly ``pasted'' on the existing label maps. Training lesion labels are mainly located in the cerebral white matter, but occurrences are also found in the cerebellum, thalamus, pallidum, and putamen. Regarding the extra-cerebral labels, these are obtained with a Bayesian segmentation approach~\citep{puonti_accurate_2020}.}
\end{figure}

\section*{Supplement 5: $\:$ Modifications to the nnUNet, and TTA, and SIFA methods}

All competing methods tested in this article are used with their default implementation, except for few minor differences that we list here. All the following modifications improve the scores obtained by the original implementations on the validation set.

\vspace{1mm}
\noindent\textbf{nnUNet~\citep{isensee_nnu-net_2021}\footnote{https://github.com/MIC-DKFZ/nnUNet}}: We now apply random flipping along the right/left axis (Dice score improvement of 0.09 on the validation set), as opposed to the original implementation where flipping was applied in any direction. This also mimics the augmentation strategy used for SynthSeg (see Section~\ref{sec:label maps}).

\vspace{1mm}
\noindent\textbf{TTA~\citep{karani_test-time_2021}\footnote{https://github.com/neerakara/test-time-adaptable-neural-networks-for-domain-generalization}}: First, the image normaliser now uses five instead of three convolutional layers, which increases its learning capacity, especially in the case of large domain gaps (Dice improvement of 0.16 on the validation set). The second modification is relative to the training atlas that is used in \cite{karani_test-time_2021} as ground-truth during the first steps of the adaptation. Here, we add an offline step, where we rigidly register this atlas to the test scan with NiftyReg (Modat et al., 2010) We emphasise that this step was not done in the original implementation, since test scans in Karani et al. were already pre-aligned. Moreover, we also increase the number of steps during which the atlas is used by increasing the beta threshold from 0.25 to 0.4~\citep{karani_test-time_2021} (Dice improvement of 0.06 on the validation set). Finally, we replace the existing data augmentation scheme by the same spatial, intensity and bias augmentations as for SynthSeg (Dice improvement of 0.03 on the validation set).

\vspace{1mm}
\noindent\textbf{SIFA~\citep{chen_synergistic_2019}\footnote{https://github.com/cchen-cc/SIFA}}: For this method, we added an online data augmentation step during training, where we apply the same spatial, intensity and bias augmentations as for SynthSeg (Dice improvement of 0.18 on the validation set).

\pagebreak

\section*{Supplement 6: $\:$ Number of retraining for each value of $N$}

\begin{table}[h]
\centering
\setlength\tabcolsep{13pt} 
    \caption{Number of label maps and associated retrainings used to assess performance against the amount of training subjects. Scores are then averaged across the retrainings. We emphasise that the number of retrainings is higher for low values of $N$ to compensate for the greater variability in random subject selection. All training label maps are randomly taken from the manual segmentations of T1-39.}
    \small
    \vspace{0.5cm}
    \begin{tabular}{l c c c c c}
        \toprule
        Number of training segmentations & 1 & 5 & 10 & 15 & 20 \\
        \midrule
        Number of retrainings  & 8 & 5 & 4 & 3 & 2  \\
        \bottomrule
    \end{tabular}
\end{table}

\section*{Supplement 7: $\:$ Training label maps for cardiac segmentation}

\begin{figure}[h]
    \centering
    \includegraphics[width=\textwidth]{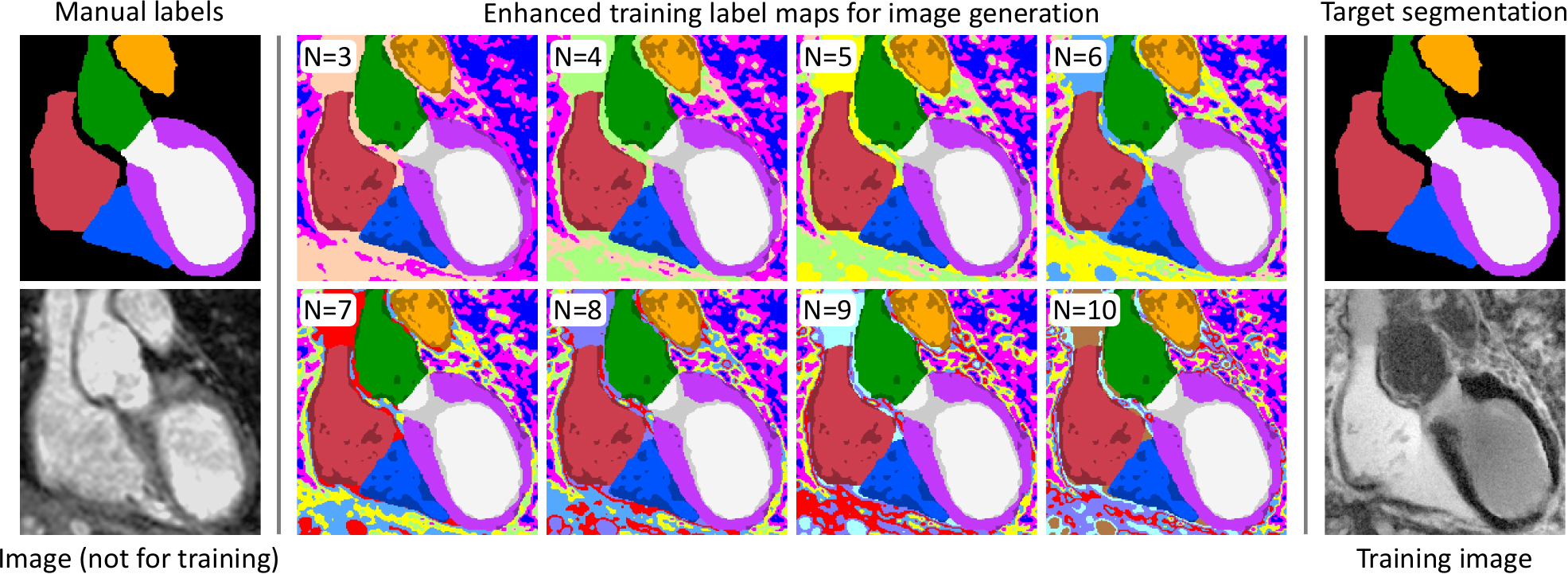}
    \caption{Training label maps for extension to cardiac segmentation are obtained by combining three types of labels. We first start with manual delineations (top left). Second, we obtain labels for sub-regions (represented in the middle label maps by slightly shaded colours) of each of these foreground regions by clustering the associated intensities in the corresponding image (bottom left). Third, we obtain automated labels for the background structures (i.e., vessels, bronchi, bones, etc., for which no manual segmentations are available), by clustering the corresponding intensities into $N$ classes ($N \in [3, 10]$), in order to model them with different levels of granularity. During training, one of these enhanced label maps is randomly selected to synthesise a training image (bottom right) by using the proposed generative model. Note that the target segmentation is reset to the initial manual labels.
    }
\end{figure}

\section*{Supplement 8: $\:$ Values of the generative model hyperparameters used in the heart experiments}

\begin{table}[h]
\centering
\setlength\tabcolsep{4pt} 
    \caption{Values of the hyperparameters controlling the generative model used in the heart experiments. As before, intensity parameters assume an input in the [0, 255] interval. Rotations are expressed in degrees, and spatial measures are in millimeters.}
    \medskip
    \small
    
    \begin{tabular}{lccccccccccccccccccc}
        \toprule
        Hyperparameter & $a_{rot}$ & $b_{rot}$ & $a_{sc}$ & $b_{sc}$ & $a_{sh}$ & $b_{sh}$ & $a_{tr}$ & $b_{tr}$ & $b_{nonlin}$ & $a_\mu$ & $b_\mu$ & $a_\sigma$ & $b_\sigma$ & $b_B$ & $\sigma_\gamma^2$ & $r_{HR}$ & $b_{res}$ & $a_\alpha$ & $b_\alpha$ \\
        \midrule
        Value          &     -45   &    45     &   0.8   &   1.2   &  -0.02  &   0.02  &   -40    &   40     &      8       &   0    &    255  &      0     &     35     &  0.7  &        0.5        &           1         &     10     &    0.95    &     1.05   \\
        \bottomrule
    \end{tabular}
\end{table}

\end{document}